\documentclass[useAMS,usenatbib]{mn2e}

\usepackage{graphicx}
\usepackage{psfrag}
\usepackage{amsmath}
\usepackage{amssymb}

\newcommand{\uJy}{\ensuremath{\mu\mathrm{Jy}}}
\newcommand{\mJy}{\ensuremath{\mathrm{mJy}}}
\newcommand{\Jy}{\ensuremath{\mathrm{Jy}}}

\newcommand{\sfind}{{\sc sfind}}
\newcommand{\imsad}{{\sc imsad}}
\newcommand{\sex}{{SE}}
\newcommand{\tesla}{{\sc Aegean}}
\newcommand{\fragnames}{\psfrag{sex}{SE}\psfrag{paaadding}{\sfind}\psfrag{selavy}{Selavy}\psfrag{imsad}{\imsad}\psfrag{tesla}{\tesla}}

\title[Compact continuum source-finding]{Compact continuum source-finding for next generation radio surveys.}
\author[Hancock et al.]{P. J. Hancock$^{1,4}$\thanks{E-mail:
Paul.Hancock@sydney.edu.au}, 
T. Murphy$^{1,2,4}$, 
B. M. Gaensler$^{1,4}$, 
A. Hopkins$^{3,4}$, \newauthor
and J. R. Curran$^{2}$\\
$^{1}$Sydney Institute for Astronomy, School of Physics A29, The University of Sydney, NSW 2006, Australia\\
$^{2}$School of Information Technologies, The University of Sydney, NSW 2006, Australia\\
$^{3}$Australian Astronomical Observatory, PO Box 296, Epping, NSW 1710, Australia \\
$^{4}$ARC Centre of Excellence for All--sky Astrophysics (CAASTRO)
}
\begin{document}

\date{Accepted 2011 ---. Received 2011 ---; in original form 2011 ---}
\pagerange{\pageref{firstpage}--\pageref{lastpage}} \pubyear{2011}
\maketitle
\label{firstpage}

\begin{abstract}
We present a detailed analysis of four of the most widely used radio
source finding packages in radio astronomy, and a program being
developed for the Australian Square Kilometer Array Pathfinder (ASKAP)
telescope. The four packages; SExtractor, \sfind{}, \imsad{} and
Selavy are shown to produce source catalogues with high completeness
and reliability. In this paper we analyse the small fraction ($\sim
1\%$) of cases in which these packages do not perform well. This small
fraction of sources will be of concern for the next generation of
radio surveys which will produce many thousands of sources on a daily
basis, in particular for blind radio transients surveys.  From our
analysis we identify the ways in which the underlying source finding
algorithms fail. We demonstrate a new source finding algorithm
\tesla{}, based on the application of a Laplacian kernel, which can
avoid these problems and can produce complete and reliable source
catalogues for the next generation of radio surveys.

\end{abstract}


\section{Introduction}
Source finding in radio astronomy is the process of finding and
characterising objects in radio images. The properties of these
objects are then extracted from the image to form a survey
catalogue. The aim of large scale radio imaging surveys is to provide
an unbiased census of the radio sky, and hence the ideal source finder
is both complete (finds all sources present in the image) and reliable
(all sources found and extracted are real).

Most of the standard source finding algorithms that have been
developed over the last few decades are highly complete and reliable,
missing only a small fraction of sources.  These problem cases are
generally dealt with in pre or post-processing, or manually corrected
in the source catalogue.

Next generation radio surveys such as the Evolutionary Map of the
Universe \citep[EMU;][]{norris_emu_2011} and the ASKAP Survey for
Variables and Slow Transients \citep[VAST;][]{chatterjee_vast_2010}
planned for the Australian SKA Pathfinder
\citep[ASKAP;][]{johnston_science_2008} Telescope will produce large area,
sensitive maps of the sky at high cadence, resulting in many times
more data than previous surveys. Data processing will need to be fully
automated, with limited scope for manual intervention and
correction. Hence the small number of missing or incorrectly
identified sources produced by current source-finders will pose a
substantial problem. In particular, in blind surveys for
radio transients, missed sources and false positives in an epoch will
cause the transient detection algorithms to trigger on false 'events'.
VAST will need to extract thousands of sources from survey images at a
cadence of $\sim5$~seconds. A source finding algorithm which is $99\%$
complete and reliable at a signal-to-noise ratio of 5, will be
producing $\sim 10,000$ false sources, and missing $\sim 10,000$ real
sources per day. Whilst it is possible to remove false sources from a
catalogue, the missing real sources are lost forever. The large data
rates of telescopes like ASKAP will make it impossible to store each
observation, and thus no reprocessing of the data will be possible.

The way in which a source finding algorithm fails to detect a real
source is often assumed to be related to noise, and that it is
random. In this paper we test this assumption and show that whilst
many sources are missed due to random noise related effects, there is
also a component that is deterministic and related to the underlying
algorithm.  By analysing the source finding algorithms and their modes
of failure we identify ways in which the algorithms could be improved
and use this knowledge to build an algorithm which can produce
catalogues that are more complete and more reliable than those
currently available.
 
In this paper we discuss the problem of source-finding for Stokes I
continuum radio emission in the context of next generation radio
imaging surveys. We will not deal directly with the additional
complications introduced by spectral line data, polarization data or
extended sources and diffuse emission.  The focus of this paper is on
upcoming surveys for ASKAP, however the results will be equally
applicable to future radio surveys on other SKA pathfinder instruments
such as the MWA \citep{lonsdale_mwa_2009} and SKAMP
\citep{adams_square_2004}, and of course the SKA itself.

In \S~\ref{sec:sf_outline} we outline the main approaches to
source-finding in radio astronomy, and then in \S~\ref{sec:alg}
describe four of most widely used source finding
packages. \S~\ref{sec:surveys} gives some examples of the way
that source finding packages are used to create catalogues for large
surveys and transient studies. \S~\ref{sec:data} describes the
test data that was used in the analysis of the source finding
algorithms, and \S~\ref{sec:eval} describes the evaluation
process. The instances in which the source finding algorithms fail to
find or properly characterise sources are described in
\S~\ref{sec:missed}. \S~\ref{sec:tesla} describes a new
source finding algorithm, \tesla{}, which has been designed to overcome
many of the problems suffered by existing source finding packages. We
summarise our conclusion in \S~\ref{sec:conclusions}.

\section{Source-finding in radio astronomy}
\label{sec:sad} \label{sec:sf_outline}
In a broad sense, source finding in radio images involves finding
pixels that contain information about an astronomical source. Most
approaches to source finding in radio astronomy follow a similar
method: (i) background estimation and subtraction; (ii) source
identification; (iii) source characterisation; and (iv) cataloguing.  In
this section we outline the standard method taken in each of these
steps.

In the discussion that follows we consider a source to be a signal of
astronomical importance that can be well modeled by an elliptical
Gaussian. By this definition a radio galaxy with a typical core/jet
morphology would be made up of three sources, one for the jet and each
of the lobes. The grouping of multiple sources into a single object of
interest (like a core/jet radio galaxy) is not in the realm of source
finding or classification as it relies on contextual information to
make such an association.

\subsection{Background estimation and subtraction}
The first step in source finding is determining which parts of the
image belong to sources and which belong to the background
\citep[eg,][]{huynh_emu_2011}.  The most common way in which this
separation is achieved is to set a flux threshold that divides pixels
in to background or source pixels. This process is referred to as
thresholding.

A straightforward case would involve a background that is dominated by
thermal noise, which is without structure and is constant across the
entire image. In such a case a single threshold value can be chosen
that will result in all sources above that threshold being detected,
and some small number of false detections. A varying background can be
accounted for by calculating the mean and rms noise in local
sub-regions, which is then used to normalize the image before applying
a uniform threshold in signal-to-noise. The selection of a threshold
limit is often a balance between detecting as many real sources as
possible and minimising the number of false detections. Typically a
$5\sigma$ threshold limit is used in a blind survey, with higher or
lower limits chosen for larger or smaller regions of sky.

False detection rate (FDR) analysis \citep{hopkins2002} determines the
threshold limit that will result in a number of falsely detected pixels
that is lower than some user defined limit.

In cases where the background has structure, an image filter must be
used to remove the background structure before the source-finding
stage. The way in which the background structure is removed depends on
the cause and type of structure that is present. A common example is
diffuse emission in the galactic plane, with compact sources embedded
within. A discussion of background filtering techniques is beyond the
scope of this paper, and in our analysis we assume the images have
been pre-processed and are free of background structure. For an
evaluation of background estimation see \citet{huynh_emu_2011}.

\subsection{Source Identification}
Source identification is the process by which pixels that are above a
given threshold are grouped into contiguous groups called
islands. Each island corresponds to one or more sources of interest.
The process of {\em finding} sources is complete at this stage.  The
format of the catalogue is just a list of pixels that belong to each
of the islands, which is not of general astronomical utility. Source
characterisation is required to convert these islands of pixels into a
more useful form.

\subsection{Source Characterisation}
Source characterisation involves measuring the properties of each
source, for example the total flux and angular size.  The best source
characterisation method is strongly dependent on the nature of the
sources that are to be studied. Point sources, by definition, have the
shape of the point-spread-function (PSF) of an image, making the PSF
shape important in the characterisation process. Images that are
produced from radio synthesis observations have been deconvolved and
the complicated PSF of the instrument has been replaced with an
appropriately scaled Gaussian. Observations with sufficient $u,v$
coverage will do not need to be deconvolved as they have a PSF that is
already nearly a Gaussian. In either case, compact sources will appear
as Gaussian, and so an island of pixels can be characterised by a set
of Gaussian components.

In lower resolution radio surveys such as the NVSS
\citep[45\,arcsec$^2$,][]{condon_nrao_1998} and SUMSS \citep[$45\times
  45{\rm cosec}|\delta|$\,arcsec$^2$,][]{mauch2003}, a majority of
objects are unresolved and can be characterised by a single
Gaussian. However, in higher resolution surveys such as FIRST
\citep[$5$\,arcsec$^2$,][]{becker_first_1995} a significant fraction
of the sources are partially extended or have multiple components, and
so multiple Gaussians are required to represent them.

Fitting a number of Gaussian components to an island of pixels is
straightforward \citep{Condon1997}, but is highly sensitive to the
choice of initial parameters. Gaussian fitting can converge to
unrealistic or non-optimal parameters due to the many local minima in
the difference function. Effective multiple Gaussian fitting requires
two things: an intelligent estimate of the starting parameters, and
sensible constraints on these parameters.  None of the widely used
source-finding packages have an algorithm for robustly estimating
initial parameters for a multiple Gaussian fit.

Two approaches have been developed which try to address the difficulty
of obtaining accurate initial parameters for multiple Gaussian
fitting: de-blending, and iterative fitting. A de-blending based
approach breaks an island into multiple sub-islands, each of which is
fit with a single component. In an iterative fitting approach, the
difference between the image data and the fitted model (the fitting
residual) is evaluated in order to determine weather an extra
component is required. This analysis will repeat until an acceptable
fit is achieved, or a limit on the number of components has been
reached. De-blending and iterative fitting are both susceptible to
source fragmentation, whereby a single true source is erroneously
represented by multiple components.

Once each island of pixels has been characterised the fitting
parameters are catalogued.

\subsection{Cataloguing}
The final stage in source finding is extracting the source parameters
and forming a catalogue of objects in the field.

A catalogue should contain an appropriate listing of every parameter
that was fit, along with the associated uncertainties. In addition to
the fitted parameters, a source finding algorithm should report
instances where the source characterisation stage was inadequate or
failed. By reporting sources that were not well fit, a catalogue can
remain complete despite having measured some source parameters
incorrectly.  Poorly fit sources can easily be re-measured, whereas
excluded sources are missed forever.  If it is not possible to
construct a reasonable facsimile of the true sky using only the
information provided in the source catalogue then the source finding
process has not been successful.

\section{Source-finding packages and their algorithms}\label{sec:alg}
Most of the major source-finding packages in astronomy are based on a
few common algorithms. In this section we outline the features of
these packages.

Source finding packages that rely on wavelet analysis were not
considered in this work as none of the most widely used source finding
packages rely on wavelet analysis.

\subsection{SExtractor}
SExtractor \citep[\sex{};][]{Bertin1996} was developed for use on
optical images from scanned plates. The speed and ease of use of
SExtractor has made it a popular choice for radio astronomy despite its
optical astronomy origins. SExtractor is a stand alone package for
Unix-like operating systems.

The source finding and characterisation process that SExtractor
follows can be modified via an extensive parameter file. For this work
the following parameters were used:
\\
\\
\begin{tabular} {ll}
\footnotesize
DETECT\_MINAREA  & 5 \\
THRESH\_TYPE     & ABSOLUTE \\
DETECT\_THRESH   & 125e-6 \\
ANALYSIS\_THRESH & 75e-6 \\
\\
MASK\_TYPE       & CORRECT \\
BACK\_SIZE       & 400 \\
BACK\_FILTERSIZE & 3 \\
\end{tabular}
\\ 

The first four parameters instruct SExtractor to detect all sources
with a peak pixel brighter than $5\sigma=125\uJy/\rm{beam}$. The source
characterisation is then carried out on islands of pixels that are
brighter than $3\sigma=75\uJy/\rm{beam}$ that contain at least 5 pixels. The
final three parameters ensure that the measured flux of a source is
corrected for the effects of nearby sources, and that the background
is estimated using a box of $3\times 400$ pixels on a side. This large
background size results in a background that is less than $1\uJy$ for
each of the tested images. The parameters DEBLEND\_NTHRESH and
DEBLEND\_MINCONT are used by \sex{} in the source characterisation
stage, when deciding how many components are contained within an
island of pixels. The ability of \sex{} to characterize sources was
found to be insensitive to these parameters for the simulated images
used in this work.

\subsection{\imsad{}} 
Image Search and Destroy (\imsad{}) is an image based source finding
algorithm in {\sc miriad} \citep{sault_retrospective_1995}. The
threshold is user--specified either as an absolute flux level or as a
signal to noise ratio (SNR) with the background noise determined from
a histogram of pixel values. Only pixels that are brighter than the
threshold are used in the fitting process. For the analysis presented
in this work we specify a threshold of $5\sigma=125\uJy/\rm{beam}$. \imsad{}
performs a single Gaussian fit to each island of pixels.

\subsection{Selavy} 
Selavy is the source finding package that is being developed by
ASKAPsoft as part of the data processing pipeline for ASKAP. Selavy is
a source finding package that is able to work with spectral cubes and
continuum images, and includes a number of different algorithms and
approaches to source finding. Selavy is related to the publicly
available Duchamp software (Whiting, 2012, MNRAS, in
press\footnote{http://www.atnf.csiro.au/people/Matthew.Whiting/Duchamp}). Selavy
is a version of the Duchamp software that has been integrated into the
ASKAPsoft architecture to run on a highly parallel system with
distributed resources. In the context of compact continuum source
finding the only difference between Selavy and Duchamp is that Selavy
is able to parameterize and island of pixels with multiple Gaussian
components. Selavy was given a threshold of $5\sigma=125\uJy/\rm{beam}$ for
source detection.

\subsection{\sfind{}}
\sfind{} \citep{hopkins2002} is implemented in MIRIAD and uses FDR
analysis to set the detection threshold. Source characterisation is
performed using the same Gaussian fitting subroutine as that use by
the MIRIAD task {\tt imfit}.

A varying background is calculated by \sfind{} by dividing the image
into sub-regions of (user defined) size, and measuring the mean and
rms of each region. In an image which contains a high density of
sources, or sub-regions which contain a particularly bright or
extended source, the calculated mean and rms will be contaminated by
the sources. For sub-regions where this occurs the result is a mean
and rms value that is significantly different from the adjacent
sub-regions, which can cause sources on the boundaries to be
normalised such that their shape and flux distribution are not
preserved. For an image which is constructed to have a zero mean and
constant rms, these contamination effects can be largely removed by
setting the size of the sub-regions to be larger than the given image.

The rejection of sources which fail to be fit with a Gaussian rejects
many instances of sources that have very few pixels. This has the
effect of further decreasing the false detection rate for the
catalogue, since a false positive source needs to have many contiguous
false positive pixels in order to be fit properly.

In this work we selected the sub-regions to be larger than the given
image, and adjusted the FDR parameter until the automatically selected
threshold was at $5\sigma=125\uJy/\rm{beam}$.

\subsection{FloodFill}
\label{sec:floodfill}
FloodFill is an algorithm which performs the second stage of source
finding, separating the foreground from the background pixels, and
grouping them into islands that are then passed on to the source
characterisation stage. We describe FloodFill as implemented in the
new source finding algorithm, \tesla{}, which is described in \S
\ref{sec:tesla}. Although used by \citet{murphy_second_2007}, the
details of the algorithm have not been described in the astronomical
literature \citep[although see][]{roerdink_watershed_2001}.

\begin{figure}
\centering
\includegraphics[width=\linewidth]{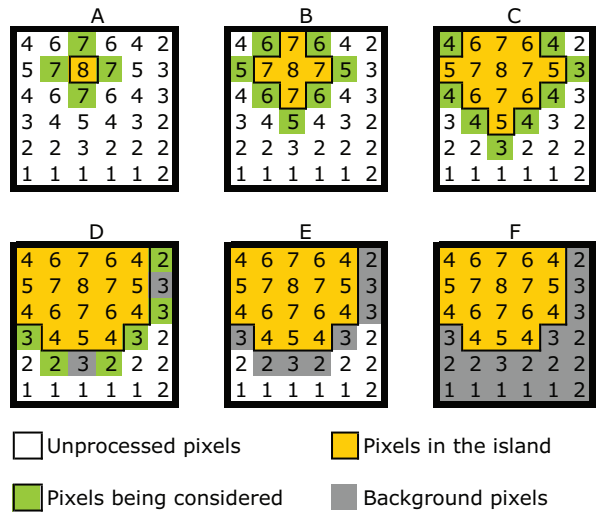}
\caption{A demonstration of the operation of the FloodFill
  algorithm. The small `image' has pixel values that are in units of
  the background noise. The seed threshold is $\sigma_s=5$ and the
  flood threshold $\sigma_f=4$. The orange pixels are those that
  have been assigned to an island, green pixels are those that are
  being considered, gray pixels have been rejected and white pixels
  are yet to be considered. In panel A the brightest pixels is used to
  seed an island. Pixels adjacent to the island are then inspected
  (coloured green). In panel B the pixels under consideration are
  brighter than the flooding threshold and are added to the
  island. The process is repeated in panels C and D. In panel E there
  are no more pixels adjacent to the island that have not been
  inspected. If panel F all pixels have either been assigned to an
  island (orange) or are labeled as background (gray).}
\label{fig:floodfill}
\end{figure}

FloodFill takes an image and two thresholds ($\sigma_s$ and
$\sigma_f$, with $\sigma_s \geq \sigma_f$). Pixels that are above the
seed threshold $\sigma_s$ are used to seed an island, whilst
pixels that are above the flood threshold $\sigma_f$ are used to
grow an island. Given a single pixel above $\sigma_s$, FloodFill
considers all the adjacent pixels. Adjacent pixels that are above
$\sigma_f$ are added to the island and pixels adjacent to these are
then considered. This iterative process is continued until all
adjacent pixels have been considered. The operation of FloodFill is
demonstrated on a simplistic `image' in Figure~\ref{fig:floodfill}. In
panel A the brightest pixel in the image has been chosen to seed the
island, and is coloured yellow. The adjacent pixels are coloured
cyan. In panel B the pixels that are adjacent to the seeding pixel are
added to the island as they are brighter than $\sigma_f=4$. Pixels
adjacent to the island are now considered. The process is repeated in
panel C. In panel D some of the adjacent pixels are now below
$\sigma_f$ and are thus not added to the island, and are flagged as
background pixels. In panel E there are no longer any pixels adjacent
to the island which have not been rejected so the search for new
pixels halts. In panel E there are no longer any pixels above the
seeding limit of $\sigma_s=5$ so all remaining pixels are flagged as
background pixels (panel F).

The operation of FloodFill is invariant to changes in the order in
which the seed pixels are chosen.  The output of FloodFill is a
disjoint list of islands, each of which contains contiguous pixels
that are above the $\sigma_f$ limit. FloodFill does not perform any
source characterisation, although it is able to report the flux of an
island of pixels by summing the pixel intensities. The fluxes that are
reported by FloodFill have a positive bias which can be corrected as
described by Hales et al. (2011, in prep).

\subsection{\tesla{}}
FloodFill forms the basis for two new source finding algorithms: BLOBCAT
(Hales et al. 2011, in prep) and \tesla{}. Both algorithms begin with a
set of islands identified by FloodFill but characterise these islands
differently. The BLOBCAT program characterises each island of pixels
without assuming a particular source structure, where as \tesla{}
assumes a compact source structure in order to fit multiple components
to each island. Here we outline the \tesla{} algorithm, with a detailed
description differed to \S~\ref{sec:tesla}.

The \tesla{} algorithm has been implemented in Python and uses
FloodFill to create a list of islands of pixels. \tesla{} was set to
use a single background threshold of $5\sigma$ to seed the islands,
and a flood threshold of $4\sigma$ to grow the islands. This threshold
was set to $5\sigma=125\uJy/\rm{beam}$ so that we are detecting
sources above an SNR of 5.

\tesla{} uses a curvature map to decide how many Gaussian components
should be fit to each island of pixels, and the initial parameters for
each component. A Python implementation of the MPFit library\footnote{
{code.google.com/p/agpy/source/browse/trunk/mpfit/mpfit.py} revision
235} is used to fit the Gaussian components with appropriate
constraints. Each island of pixels is thus characterised by at least
one Gaussian component.

\subsection{Source finding in radio surveys}
\label{sec:surveys}
The process of creating a catalogue of sources from survey images
involves more than running one of the source finding packages
described in \S~\ref{sec:alg}. Since the observing strategy,
hardware and data reduction techniques can vary widely between
surveys, standard source finding packages are typically used only as a
{\em starting point} for the creation of a source catalogue. The time
required to carry out the observations for a large area sky survey is
typically spread over multiple years. This prolonged observing
schedule is usually accompanied by multiple iterations of calibration,
data reduction and source detection, so that by the time the final
observations are complete it is possible to produce a survey catalogue
using a source finding pipeline that has been refined over many
years. 

The NRAO Very Large Array (VLA) Sky Survey \citep[NVSS,][]{condon_nrao_1998}, drew upon
observations from 1993-1996, during which time the AIPS source finding
routine SAD was modified to create VSAD. The survey strategy for the
NVSS was devised to give noise and sidelobe characteristics that
were both low, and consistent across the sky. The configuration of the
VLA was varied across the sky to ensure a consistent resolution
throughout the survey. The survey strategy was thus designed to
produce images that were nearly uniform across the sky, making the
task of source finding as easy as possible. The completeness and
reliability of the NVSS catalogue was improved in the years subsequent
to the completion of the survey with the final stable release in 2002.

The Sydney University Molonglo Sky Survey \citep[SUMSS,][]{mauch2003}
and Molonglo Galactic Plane Survey
\citep[MGPS-2,][]{murphy_second_2007} both used the source finding
package VSAD, however the single purely East--West configuration of
the telescope meant that the resolution varied with declination, and
the regularly spaced feeds produced many image artefacts. The changing
resolution and image artefacts meant that the source finding algorithm
produced many false detections. The image artefacts appeared as radial
spokes or arcs around bright sources. In order to rid the source
catalogue of falsely detected sources, a machine learning algorithm
was implemented \citep{mauch2003,murphy_second_2007}. The machine
learning algorithm was able to discriminate between real and false
sources, but required substantial training to achieve high
completeness and reliability.

An archival transients survey has recently been completed using the
data from the SUMSS survey \citep{bannister_2011}. In an archival
search spanning 20 years of observations, the need for a fast source
detection pipeline is not important, as fast transients will not be
detected, and slow transients will remain visible for years to
come. In the \citet{bannister_2011} survey, regions of sky with
multiple observations were extracted from the archival SUMSS
data. These regions of sky were analysed for sources which either
changed significantly in flux, or which were detected in only a subset
of the images. The \sfind{} package was used to detect sources, which
were then remeasured using the {\sc miriad} routine {\sc imfit}. A
complication that was encountered in the analysis of the SUMSS data,
was the contamination of candidate source lists due to source finding
errors. False positive detections and missed real sources both appear
as sources which are only detected in a subset of all the images, and
thus appear to be transient sources. The light curve of each transient
event therefore needed to be double checked in order to remove such
occurrences.

A similar transient detection project was carried out with new
observations from the Allen Telescope Array, in the ATA Transients
Survey \citep[ATATS,][]{croft_2011}. Side--lobe contamination in the
ATA images is much lower than that in the SUMSS images used in the
\citet{bannister_2011} study, however falsely detected sources in the
individual images still resulted in false transient detections and
required further processing to remove.

The process of finding sources and creating a catalogue extends beyond
the operation of a source finding package and has previously required
substantial manual intervention. The next generation of telescopes,
particularly the dedicated survey instruments, will be able to
complete observations on a much shorter time scale than current
generation telescopes, and thus the time spent creating the refining
the source catalogue will become a larger fraction of the total
effective survey time. Source finding packages that are able to
produce more accurate, complete and reliable catalogues will provide a
better starting point for the final version of the survey catalogue.

\section{Test Data}
\label{sec:data}
We used a simulated data set to evaluate the source-finding algorithms
described in \S~\ref{sec:alg}. A simulated data set has the
advantage that we are able to control the image properties (such as
rms noise) and that we know the input catalogue.

Matching recovered sources with a true list of expected sources is an
important part of the analysis presented in this paper. With any real
data set, the list of expected sources comes with some degree of
uncertainty, in that these lists are recovered from incomplete and
noisy reconstructions of the radio sky. To avoid such uncertainties we
generated a master catalogue of sources, which was then used to create
a simulated image of the sky. With absolute control over the input
catalogue and image characteristics, we are able to make more
definitive statements about the quality of the catalogues that are
produced.

\begin{figure*}
\centering
\includegraphics[width=\linewidth,bb=0 0 512 512,clip=]{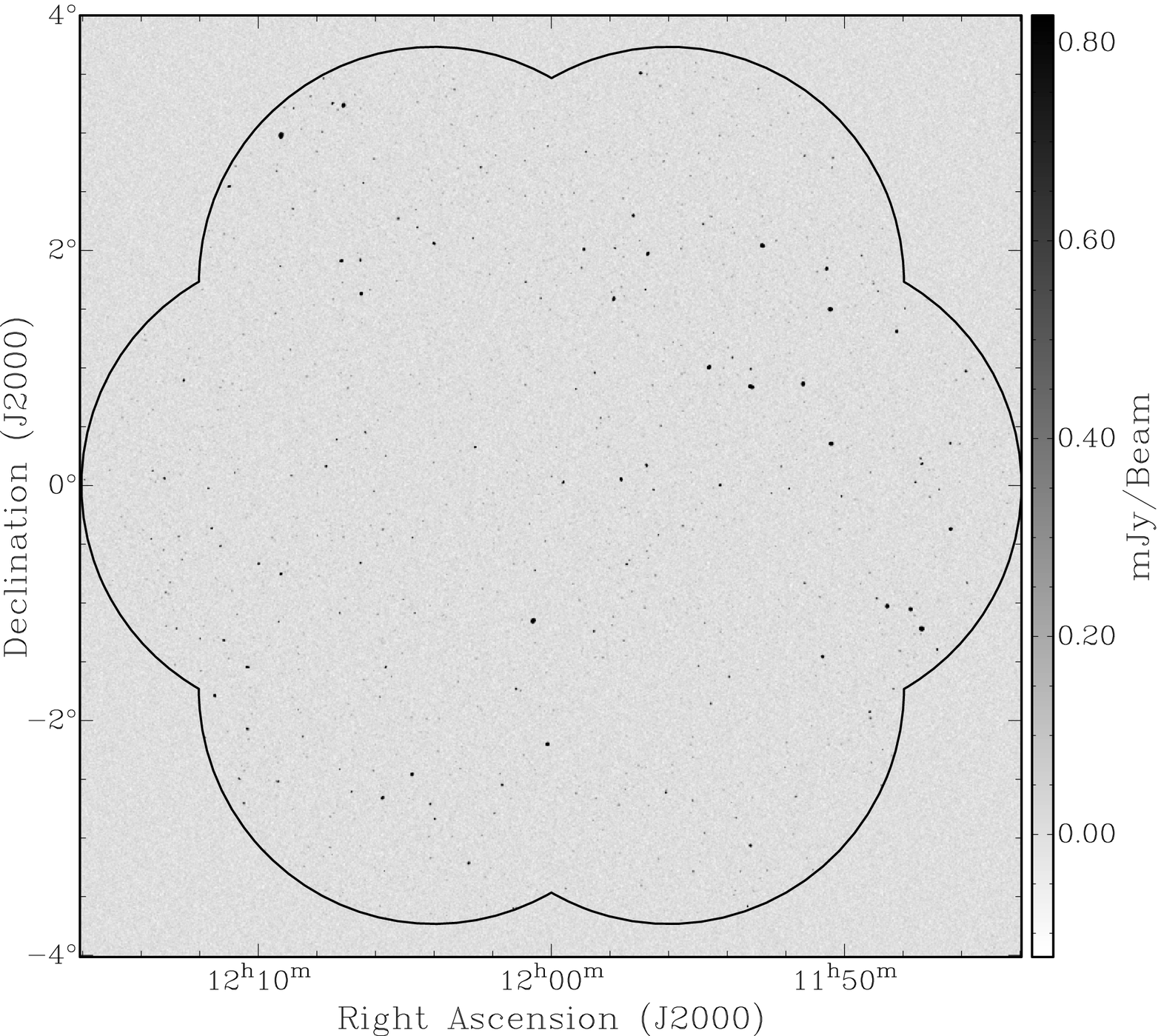}
\caption{Simulated image of the sky. The black line delineates the region of sky containing the injected sources. The colour bar ranges from $-5\sigma$ to $33\sigma$.}
\label{fig:sim_sky}
\end{figure*}

The master source catalogue was generated with the following
constraints:
\begin{enumerate}
\item Fluxes: The source peak flux are distributed as $N(S)\propto S^{-2.3}$, and within the range $(25~\uJy, 10~\Jy)$.
\item Positions: Sources were randomly distributed in space within one of ten regions of sky similar to that in Figure~\ref{fig:sim_sky}. Source clustering was not considered.
\item Morphologies: The major and minor axes of each source were randomly distributed in the range $0-52\arcsec$ with position angles in the range $(-90\deg, +90\deg)$.
\end{enumerate}

A simulated sky image was created, which contained each of the sources
in the input catalogue. The image has a 30\arcsec\ sythesised beam,
and a $25\uJy/\rm{beam}$ rms Gaussian background noise. The sources were
injected with a peak flux and morphology as listed in the
catalogue. The size of the image is 4801x4801 pixels with a scale of
6\arcsec\ per pixel, resulting in a synthesised beam sampling of 5
pixels per beam. Regions of sky exterior to the catalogue contain
noise but no sources.

The simulated data-set can be found online at
\verb+www.physics.usyd.edu.au/~hancock/simulations+

\section{Source-Finding Evaluation}
\label{sec:eval}
The source finding packages described in \S~\ref{sec:alg} were used to
generate a catalogue of sources from the simulated images. Each source
finding package was run with a $5\sigma$ threshold. In the case of
\sfind{}, the FDR was chosen so that the resulting threshold was equal
to $5\sigma$.

The source finding algorithms were evaluated by comparing these
catalogues with the input source catalogue. Three standard metrics
that have been used in the comparison of catalogues, and hence source
finders, are the completeness, reliability, and flux distribution, as
defined and discussed in \ref{sec:fluxes} - \ref{sec:correctness}
below.

\subsection{Cross matching of catalogues}\label{sec:crossmatching}
Much of the analysis that will be discussed in
\S~\ref{sec:fluxes}-\ref{sec:correctness} relies on the
cross--identification of sources from two catalogues. A common
criterion for accepting cross--identifications between catalogues is
to choose the association with the smallest sky separation, up to a
maximum matching radius.  To decrease the chances of false
associations we also consider the flux of the source when choosing
between multiple matches within a matching radius of 30\arcsec. The
distance in phase space, $D$, is given by:
\begin{equation}
-\log(D) = \frac{(\alpha_1-\alpha_2)^2}{\sigma^2_\alpha} + \frac{(\delta_1-\delta_2)^2}{\sigma^2_\delta} + \frac{(S_1-S_2)^2}{\sigma^2_S}
\end{equation}
where $(\alpha,\delta)$ are (RA,DEC), $S$ is the flux, and
$\sigma_\alpha=\sigma_\delta=30\arcsec$ is the size of the convolving
beam, and $\sigma_S=25\uJy/\rm{beam}$ is the image rms noise.

\subsection{Flux Distribution}\label{sec:fluxes}
\begin{figure}
\fragnames
\centering
\includegraphics[height=\linewidth,bb=105 145 501 655,angle=-90]{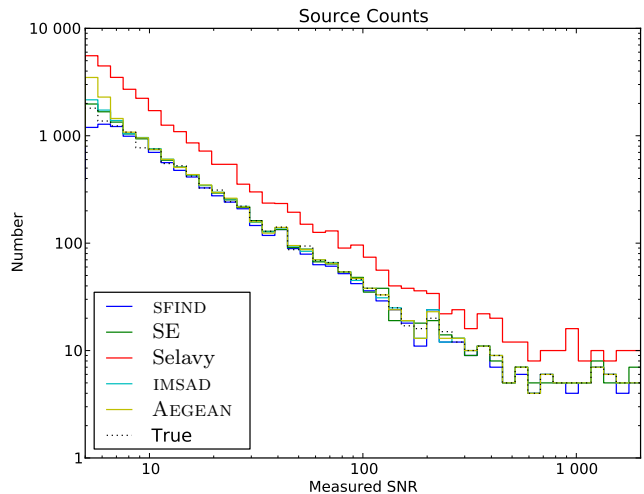}
\caption{The source count distribution of the catalogued sources. The true source count distribution is shown as a dotted line. Selavy consistently reports twice the true number of sources.}
\label{fig:flux_dist}
\end{figure}

The analysis of the flux distribution of a catalogue does not require
catalogues to be cross-matched. Since the input source catalogue was
constructed with a particular flux distribution, we should expect to
see this distribution replicated in the output
catalogues. Figure~\ref{fig:flux_dist} shows the flux distribution for
each of the source finders compared to the input distribution. Except
for Selavy, each of the catalogues have a flux distribution that is
consistent with the input catalogue. An excess of sources can be a
sign of spurious detections, whilst a lack of sources can be due to
incompleteness. If a source finding algorithm has a flux distribution
that deviates from the ideal case, it indicates that something is
wrong however the cause of the problem cannot be identified from this
graph alone. Selavy has around double the number of sources at all
flux levels as it suffers from source fragmentation. Since the
fragmented components are close to the true position of the original
source, the completeness and reliability of Selavy are only partly
compromised.

\subsection{Completeness}\label{sec:completeness}
The completeness of a catalogue at a flux $S_0$ is often defined as
the fraction of real sources with true flux $S\geq S_0$ that are
contained within the catalogue. In practice the completeness is
measured as the number of sources with a {\em measured} flux $S\geq
S_0$ that are contained within the catalogue. The two measures are
comparable at large SNR, but when the SNR is $\sim5$ the flux of a
source can be in error by $\sim 20\%$. The completeness relative to
the {\em measured} source fluxes is also effected by
\citet{eddington_1913} bias, whereas the completeness relative to the
{\em true} source fluxes is not.

\begin{figure}
\fragnames
\centering
\includegraphics[width=\linewidth,bb=50 190 550 605]{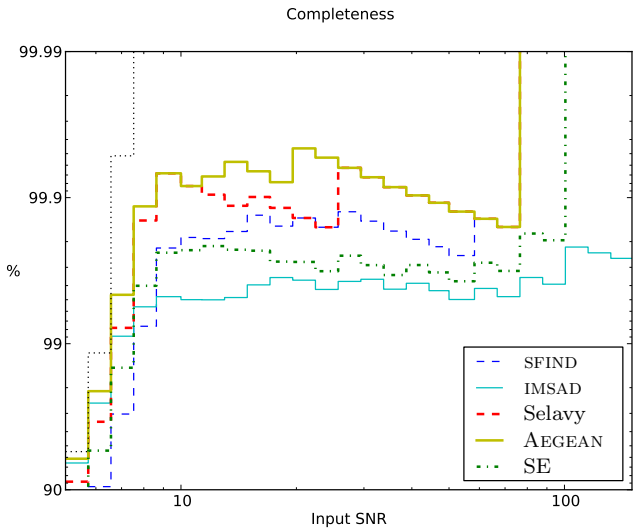}
\caption{The completeness of each catalogue as compared to the input
  catalogue. The coloured curves represent the completeness of the
  named source finders. The black dotted curve represents the expected
  completeness of an ideal source finder as calculated by
  Equations~\ref{eq:theoretical_completeness}-\ref{eq:flux_probability}.}
\label{fig:completeness}
\end{figure}

The completeness of a source finder was determined by matching the
simulated catalogue with each of the source finding catalogues. The
completeness of a catalogue at a flux $S_0$ is then the fraction of
real sources of flux greater than $S_0$ which are contained within
the given catalogue. Figure~\ref{fig:completeness} shows completeness
as a function of injected SNR for each of the source finders. Plotted
alongside each of the completeness curves is a theoretical expectation
of completeness for comparison. The expected completeness has been
determined by taking each of the sources in the input catalogue and
calculating the probability that it will be seen at a particular
flux level, given the known rms in the image. The expected
completeness is calculated as

\begin{align}
\mathcal{C}(S_0) &= \frac{\sum_{S}P(S>S_0)}{\sum_{S>S_0}N(S)}\label{eq:theoretical_completeness}\\
P(S>S_0) &= \frac{1}{\sigma}\sqrt{\frac{4\log2}{\pi}}\int_{S_0}^\infty e^{-4\log2\left(\frac{S^\prime -S}{\sigma}\right)^2}dS^\prime \label{eq:flux_probability}\\ 
 &= \frac{1}{2} \mathrm{Erfc} \left(\frac{ \sqrt{4\log2}(S_0-S)}{\sigma} \right) \notag
\end{align}

where $P(S>S_0)$ is the probability that a source of flux $S$ will be
seen at a flux greater than $S_0$ after noise has been included, and
$N(S)dS$ is the number of sources with flux between $S$ and
$S+dS$. Erfc is the complementary error function. At SNR $> 6$ all of
the source finding packages produce catalogues that are greater than
$99\%$ complete. The high completeness is due to, and responsible for,
the wide--spread use of the given source finders. The different levels
of completeness shown in Figure~\ref{fig:completeness} is a direct
result of the source finding algorithms implemented by each of the
packages. The performance of \sfind{} is comparable to the other source
finders above an SNR of 7, but less complete below this SNR. The lower
completeness is a result of \sfind{} 's focus on minimising the false
detection rate, as is shown in Figure~\ref{fig:fdr}. Selavy and \tesla{}
are the most complete source finding packages at all SNRs, however
Selavy achieves this at a cost of an increased false detection rate
(see \S~\ref{sec:fdr} and Figure~\ref{fig:completeness}). \tesla{}
is able to achieve high completeness and low false detection rate at
all SNRs. The completeness of each of the source finding packages is
summarised in Table~\ref{tab:cr_stats}.

\subsection{False Detection Rate}\label{sec:fdr}
\begin{figure}
\fragnames
\centering
\includegraphics[width=\linewidth,bb=50 190 550 605]{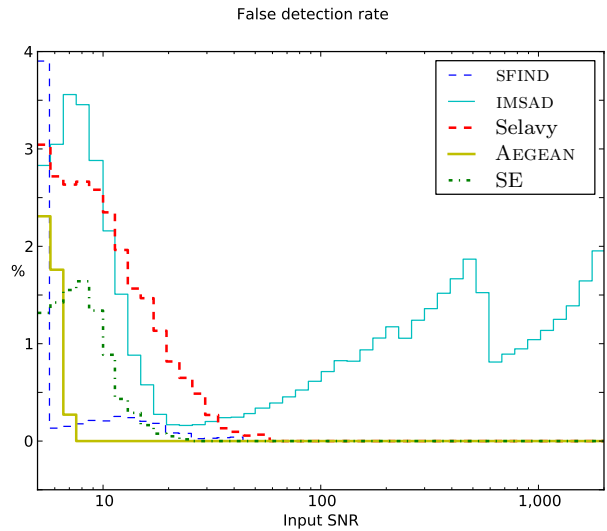}
\caption{The false detection rate (FDR) for each of the source finding
algorithms. The FDR is entirely a function of the source finding
algorithm. No falsely detected sources are expected above an SNR of 5
for the area of sky simulated.}
\label{fig:fdr}
\end{figure}

The false detection rate (FDR) of a source finding package at a flux
$S_0$, is defined as the fraction of catalogued sources with $S\geq
S_0$ which are not identified with a real source. The FDR of a source
finder was determined by matching the resulting catalogue with the
simulated source list. Catalogued sources which are not within
$30\arcsec$ of a true sources are considered false detections. The
false detection rate of a source finding algorithm is related to the
commonly used metric of reliability by:
\begin{equation}
\mathrm{FDR} + \mathrm{Reliability} = 100\%
\end{equation}
In Figure~\ref{fig:fdr} the FDR is plotted as a function of SNR for
each of the source finding packages. Substituting a flux of $S=0\,uJy$
into equation~\ref{eq:flux_probability} and considering the area of
sky covered by the simulated images we expect that there is less than
one false detection due to random chance. Thus an ideal source finding
algorithm should have an FDR of zero. The $>1\%$ FDR peaks shown in
Figure~\ref{fig:fdr} (especially for \imsad{}) are due to islands of
multiple sources that have not been properly
characterised. SExtractor, \imsad{} and Selavy have a higher false
detection rate than \sfind{} and \tesla{}, as the former are not able
to accurately characterise islands of pixels.

The single sources that are fragmented into multiple sources by Selavy
often have positions that are close enough to the true position that
they are not considered false detections, and thus don't significantly
impact the false detection rate. However at low SNRs, Selavy breaks
single sources into three or even four components, and one or more of
these components have a position distant enough from the true source
that they are registered as false detections. This is evident in
Figure~\ref{fig:fdr}.

\imsad{} suffers from the reverse problem to Selavy, in that it will
never break islands into multiple components even when they contain
multiple sources. The position that is reported by \imsad{} in such
situations can be sufficiently far from the true position that these
islands are registered as false detections. All of the false
detections for \imsad{} above an SNR of 20 in Figure~\ref{fig:fdr} are
due to this flaw.

The reliability of each of the source finding packages is summarised
in Table~\ref{tab:cr_stats}.

\begin{table}
  \begin{tabular}{l | c c c | c c c }
Package    & \multicolumn{3}{|c|}{Completeness (\%)}& \multicolumn{3}{|c|}{Reliability (\%)}\\  
           & $5\sigma$ & $10\sigma$ & $50\sigma$    & $5\sigma$ & $10\sigma$ & $50\sigma$ \\    
\hline     					                                                
\imsad{}	   & 93.44     & 99.50      & 99.49 	    &  97.17    & 97.75      & 99.66 \\
Selavy	   & 91.20     & 99.92      & 99.87 	    &  96.96    & 97.65      & 99.93 \\
\sex{}	   & 88.31     & 99.77      & 99.62 	    &  98.68    & 99.11      & 100.0 \\
\sfind{}	   & 82.48     & 99.81      & 99.75 	    &  96.09    & 99.79      & 100.0 \\
\tesla{}     & 93.87     & 99.91      & 99.87 	    &  98.69    & 100.0      & 100.0 \\
Ideal      & 94.51     & 100.0      & 100.0 	    &  100.0    & 100.0      & 100.0 \\
\hline
  \end{tabular}
\caption{The completeness and reliability of each of the source finding algorithms. The 'Ideal' case has been included for comparison.}
\label{tab:cr_stats}
\end{table}

\subsection{Measured parameter correctness}\label{sec:correctness}
For all measured catalogue sources that were identified with a true
source it is possible to compare the measured parameters to the known
true values.

\begin{figure}
\fragnames
\centering
\includegraphics[height=\linewidth, bb=120 160 500 655, angle=-90]{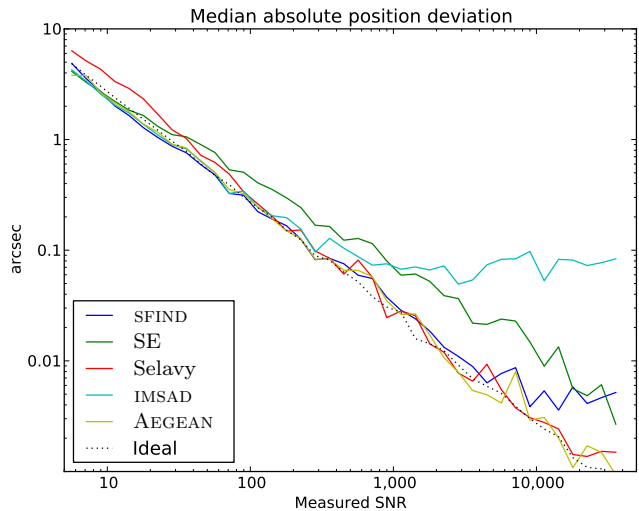}
\caption{The accuracy with which each of the source finding packages
determines the position of sources. The dotted gray curve is the
expected accuracy for an ideal elliptical Gaussian fit.}
\label{fig:positions}
\end{figure}

Figure~\ref{fig:positions} shows the median absolute deviation (MAD)
in position, as a function of SNR, for each of the source finding
algorithms. The MAD is calculated for each SNR bin and is not a
cumulative measure. An ideal source finding algorithm will have a
typical error in position that is proportional to $C/\mathrm{SNR}^2$
where $C$ is a constant that depends on the morphology of the source
and the convolving beam (see \citet{Condon1997} for detailed a
calculation). The MAD in position of an ideal source finding algorithm
is calculated semi-analytically by assuming that each source in the
input catalog has measurement error of $C/\mathrm{SNR}^2$. This ideal
curve is plotted in Figure~\ref{fig:positions}.

As is expected, the accuracy with which a source position can be
measured increases with flux, and is in agreement with the performance
of an ideal Gaussian fitting routine, which is shown as a dotted curve
in Figure~\ref{fig:positions}. The deviations from ideal behaviour
that can be seen in Figure~\ref{fig:positions} for the various source
finders at high SNR are artifacts of the reporting accuracy of the
packages. For example, \imsad{} reports positions to a resolution of
0.1\arcsec\ and therefore cannot achieve a median absolute deviation
in position better than $\sim 0.1\arcsec$. \sfind{} has similar problems
at an SNR of $\gtrsim 3000$. The median absolute position deviation
for \tesla{} and Selavy will also deviate from ideal, but at an SNR in
excess of the $40,000$ reported in
Figure~\ref{fig:positions}. SExtractor does not use Gaussian fitting
to characterise source positions and therefore does not perform as
well as the ideal at SNR greater than 50. At an SNR of $<100$ Selavy
has a median absolute position deviation that is higher than the
ideal. This is because of source fragmentation.

\begin{figure}
\fragnames
\centering
\includegraphics[height=\linewidth, bb=120 160 500 655, angle=-90]{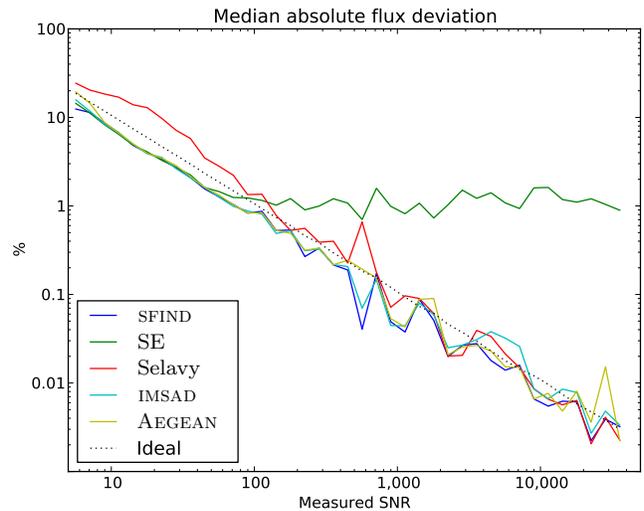}
\caption{The accuracy with which each of the source finding packages
measures the flux of sources as a function of the reported flux. The
dotted gray line is the expected accuracy for an ideal elliptical
Gaussian fit.}
\label{fig:fluxes}
\end{figure}

Figure~\ref{fig:fluxes} shows the MAD in flux as a fraction of total
flux, as a function of SNR.  Again the ideal behavior of a Gaussian
fitter has been shown by a dashed curve. Overall the source finding
packages report fluxes that are consistent with the expected ideal
Gaussian fit, the exceptions being SExtractor above an SNR of 50, and
Selavy at an SNR below 50. SExtractor deviates from the ideal and has
a plateau at $1\%$ flux accuracy. In this work we use the corrected
isophotal fluxes (FLUX\_ISOCOR) from SExtractor. Of all the methods
that are available for measuring fluxes in radio synthesis images, the
corrected isophotal fluxes was found to be the most accurate. Selavy
deviates from the ideal case and has a flux accuracy of about $1/2$ of
ideal. The cause of this deviation is source fragmentation in which
each component has only a fraction of the total true flux (see
\S~\ref{sec:fluxes} and Figure~\ref{fig:flux_dist}).

\subsection{Initial evaluation summary}
Each of the source finding algorithms conforms to a high standard of
completeness and reliability, and is able to produce a robust
catalogue of a statistically large number of sources, with accurate
measurements of position and flux. The completeness of the \tesla{}
source finding package is as good or better than any of the other
packages, and has been achieved without sacrificing reliability. In
the context of next generation radio surveys, we are interested in the
small differences between each of these source finders, and in how to
optimize the approach to avoid even the residual small level of
incompleteness and false detection rate. In surveys such as EMU
\citep{norris_emu_2011}, with an expected 70 million sources, an FDR
of even 1\% translates into $700,000$ false sources. This clearly has
an impact on the study of rare or unusual behaviour. In particular we
are interested in how the source finding algorithm affects the final
output catalogue at a level that is far more detailed than previously
explored. With this in mind we now delve into specific cases in which
existing source finding packages fail.

\section{Missed sources}
\label{sec:missed}
We are now at the stage where we can consider the real sources that
were missed by the source finding packages, as well as the false
detections that these programs generate. There are two populations of
sources that are missed by one or more of the source finding packages
as will be discussed in \S~\ref{sec:isolated}-\ref{sec:multiple}
below.

\subsection{Isolated faint sources}\label{sec:isolated}
In the simulated image, for which no clustering was taken into account,
$99.5\%$ of the islands contained a single source.

The first population of sources that was not well detected by the
source finding algorithms are isolated faint sources. These sources
have a true flux greater than the threshold, but have few or no pixels
above the threshold due to the addition of noise. \sfind{} and
SExtractor require an island to have more than some minimum number of
pixels for it to be considered a candidate source. \imsad{}, Selavy and
\tesla{} have no such requirement. The number of sources that are not
seen in a catalogue due to the effects of noise can be calculated
directly and is essentially the inverse problem to that of false
detections. A correction can be applied to any statistical measure
extracted from the catalogue in order to account for these missed
sources. The only way to recover all sources with a true flux above a
given limit, is to have a threshold that is well below this limit,
either by producing a more sensitive image, or by accepting a larger
number of false detections. Since this noise affected population of
sources cannot be reduced by an improved source finding algorithm, and
can be accounted for in a statistically robust way, we will consider
this population to be non-problematic.

\subsection{Islands with multiple sources}\label{sec:multiple}
\begin{figure}
\centering
\begin{tabular}{cc}
Image & \sfind{} \\
\includegraphics[width=0.4\linewidth]{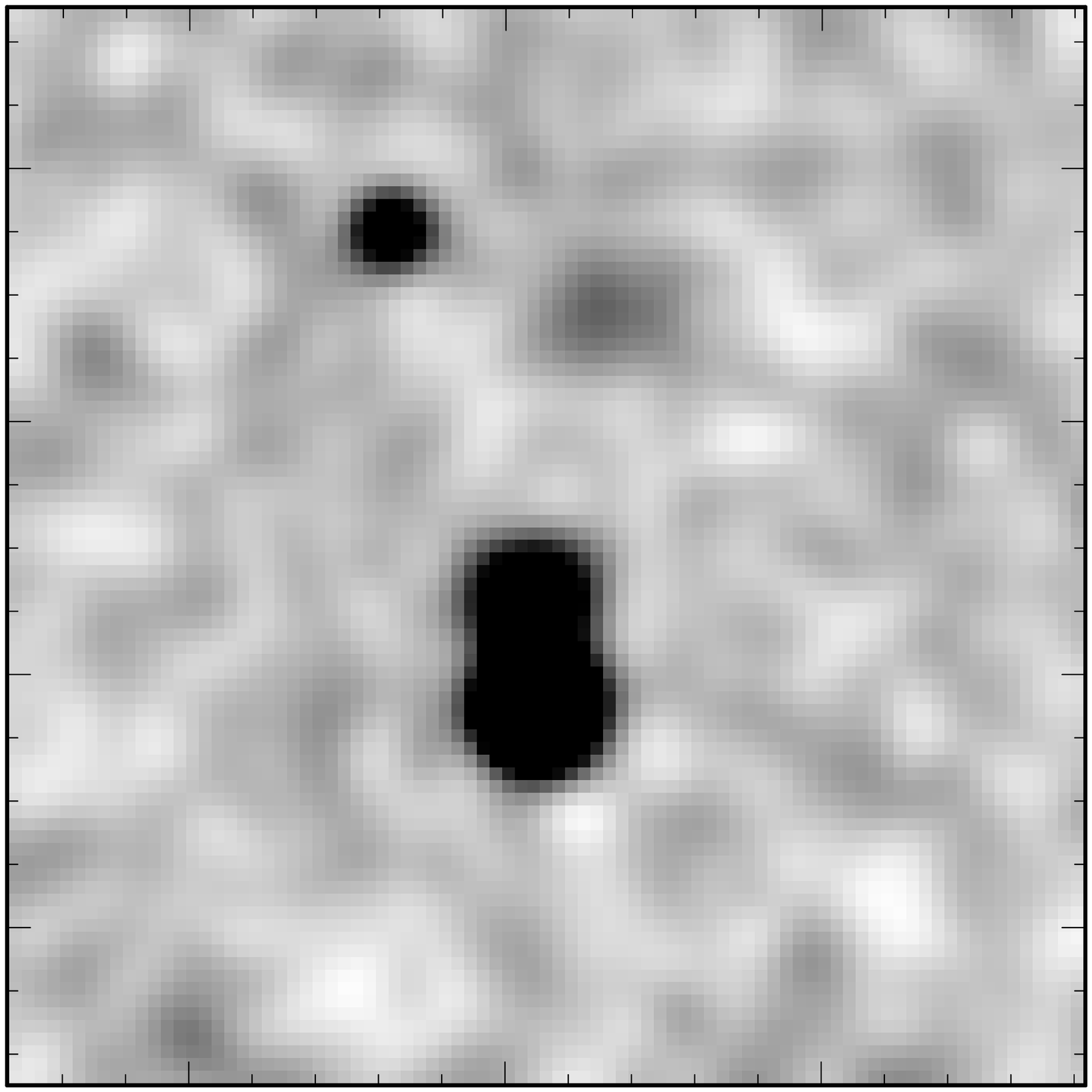} &
\includegraphics[width=0.4\linewidth]{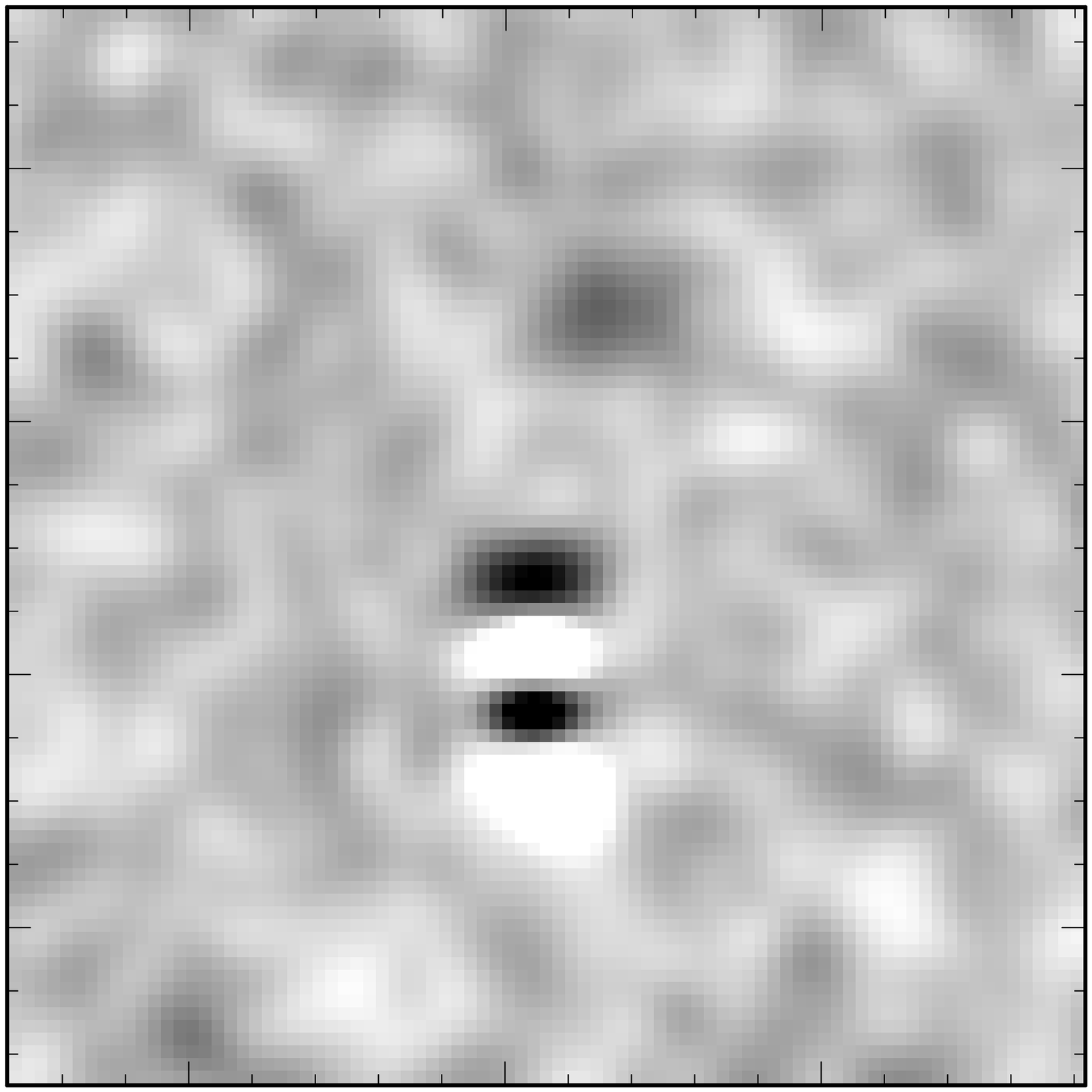} \\
\sex{} & Selavy\\
\includegraphics[width=0.4\linewidth]{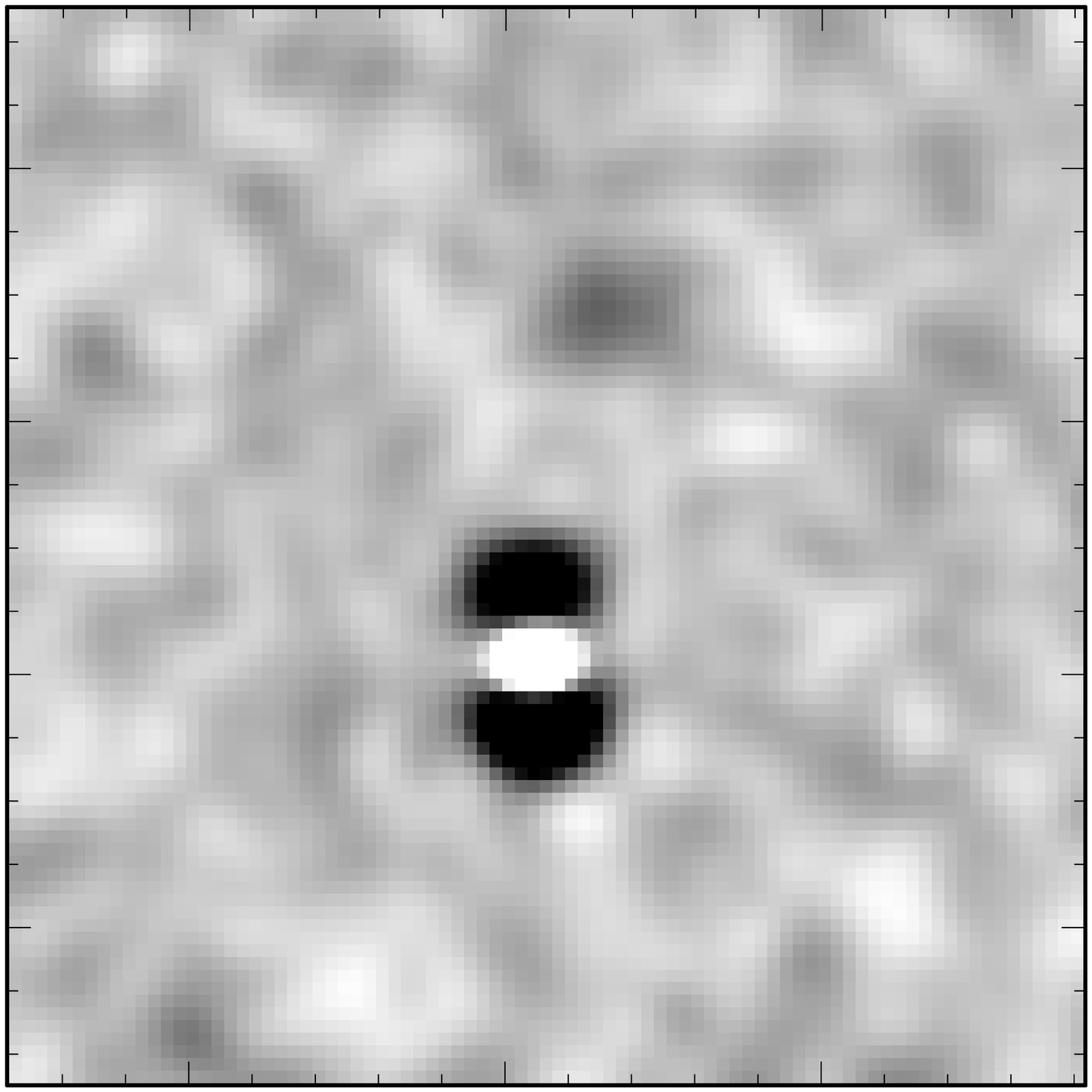} & 
\includegraphics[width=0.4\linewidth]{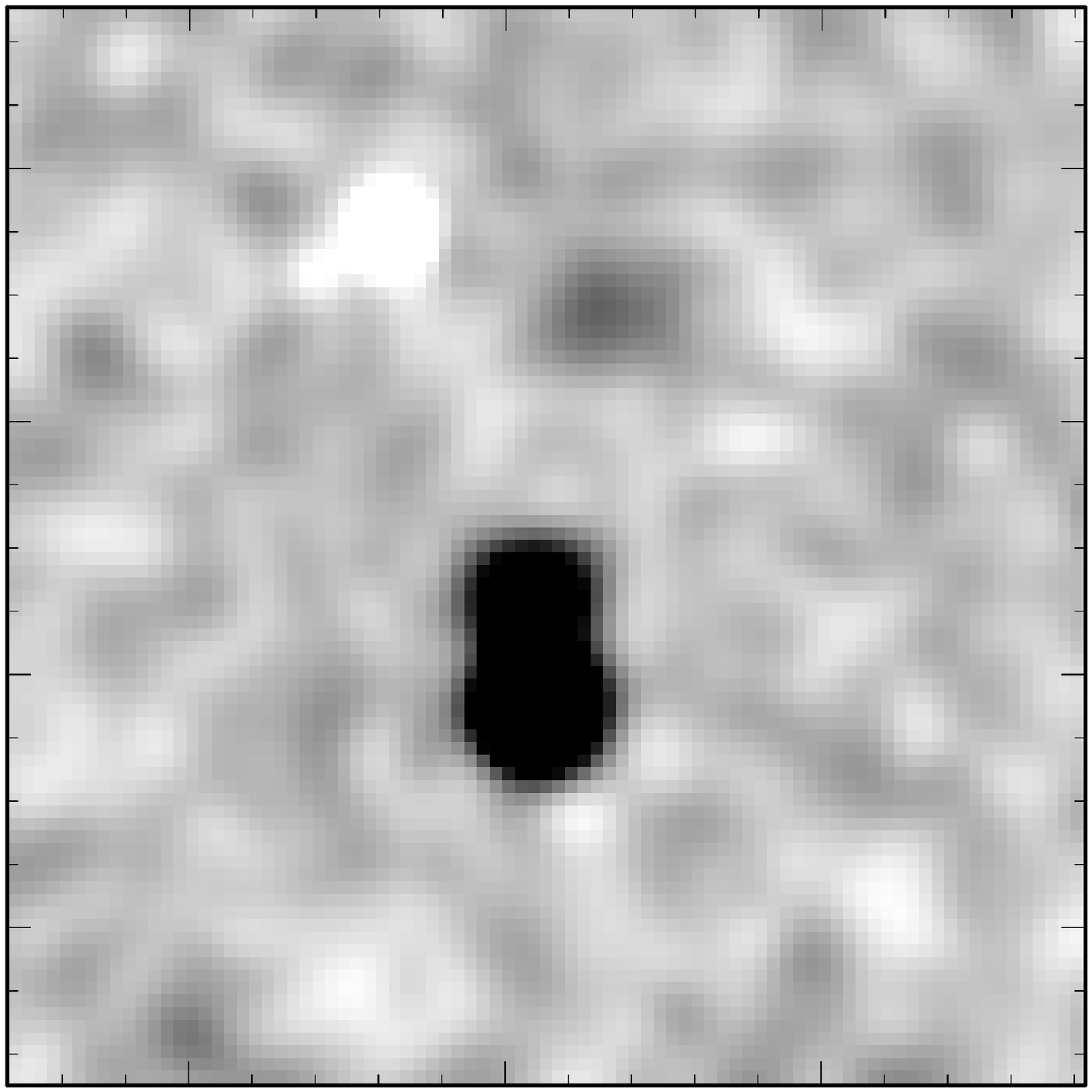} \\
\imsad{} & \tesla{} \\
\includegraphics[width=0.4\linewidth]{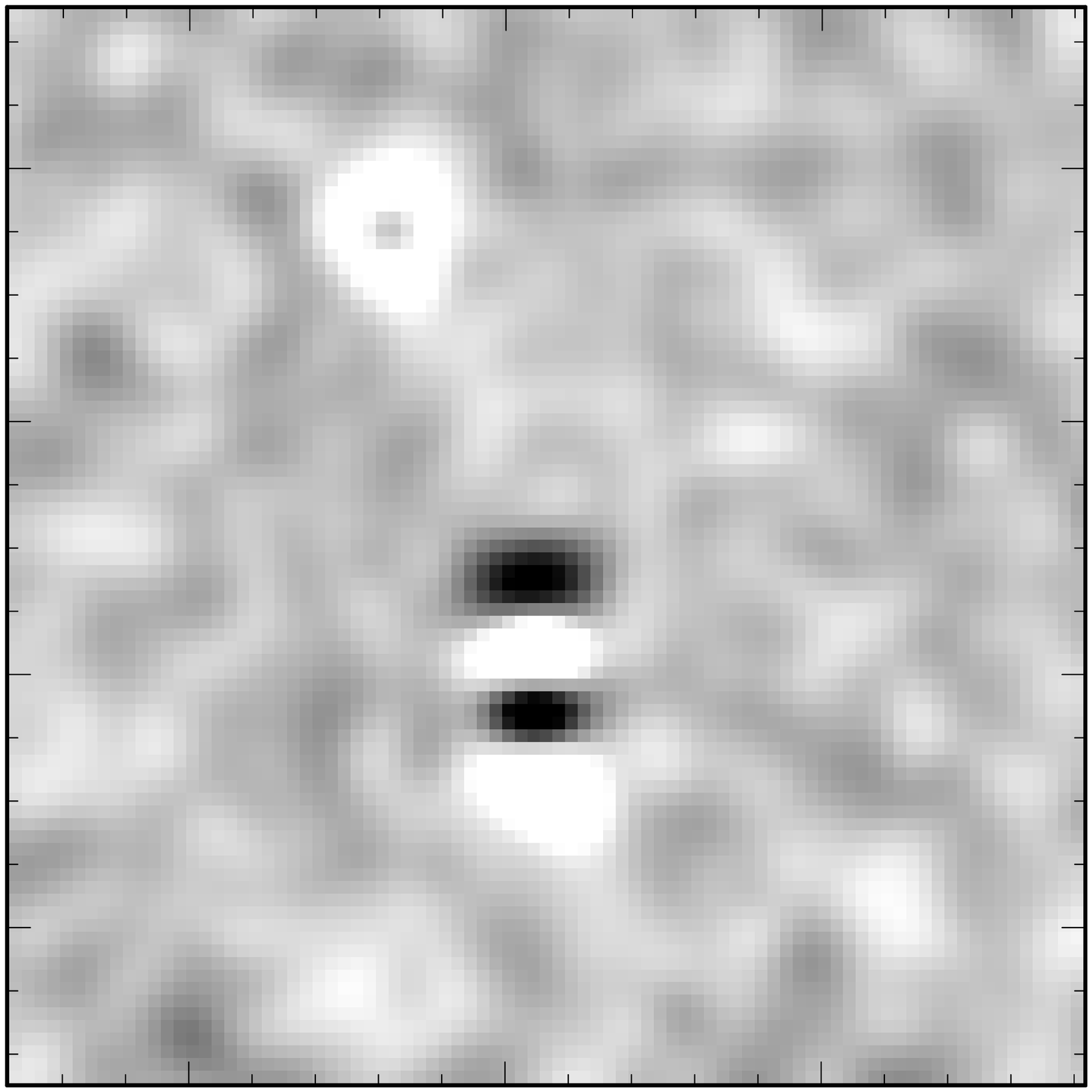} &
\includegraphics[width=0.4\linewidth]{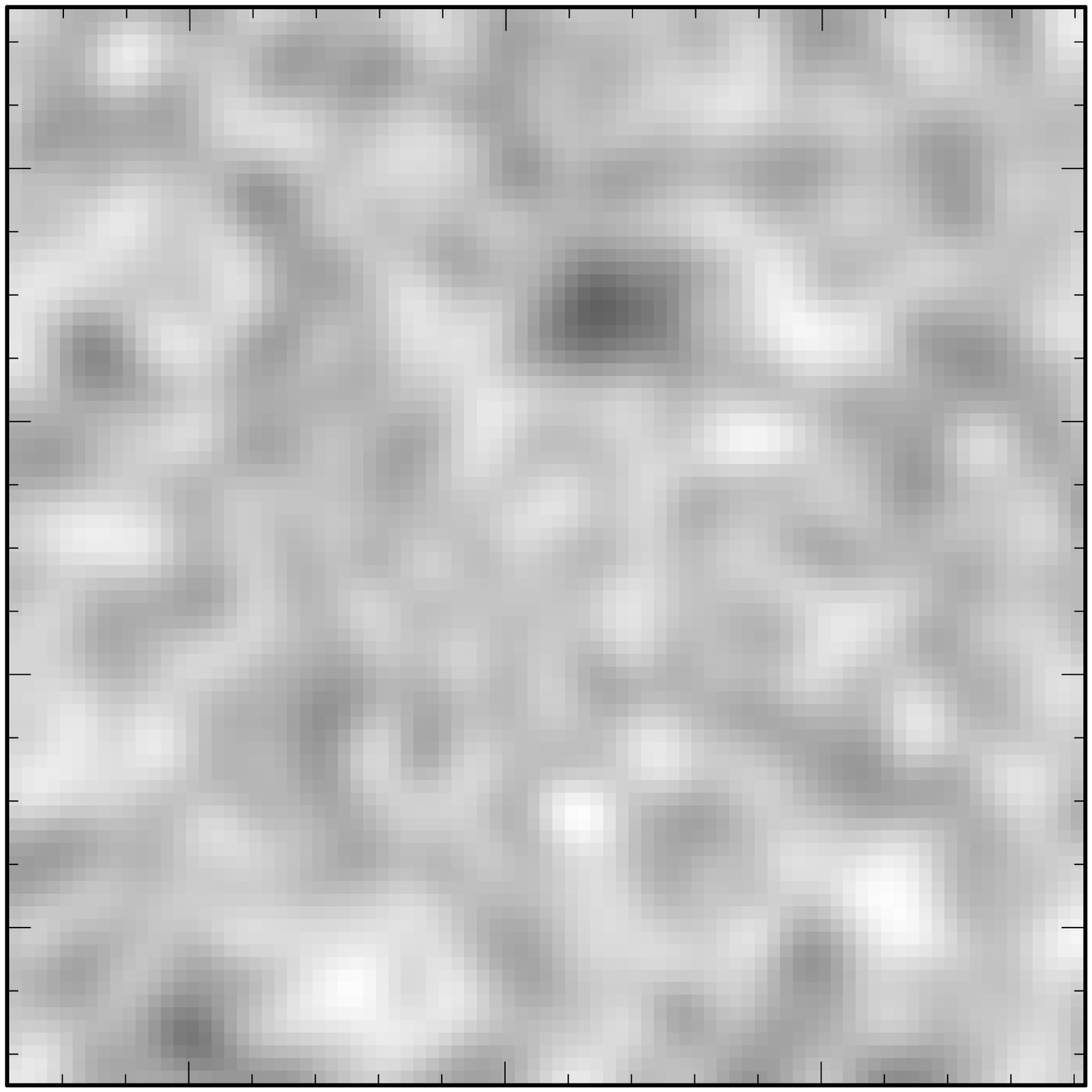} \\
\end{tabular}
\caption{{\em Top Left}: A section of the simulated image. {\em Remainder}: The fitting residual for each of the source finding algorithms. \tesla{} was the only algorithm to fit all three sources, over both islands.}
\label{fig:residual_196}
\end{figure}

\begin{figure}
\centering
\begin{tabular}{cc}
Image & \sfind{} \\
\includegraphics[width=0.4\linewidth]{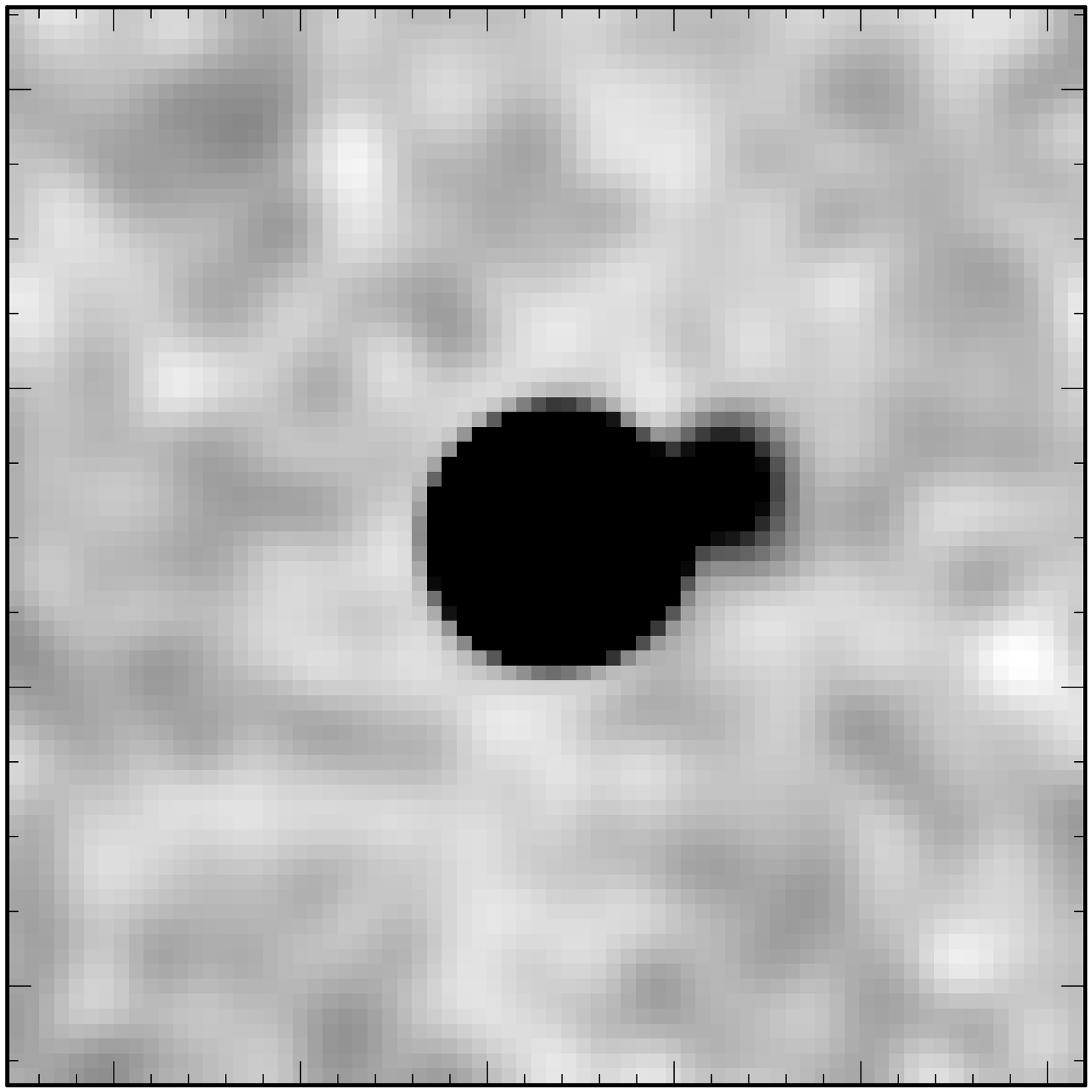} &
\includegraphics[width=0.4\linewidth]{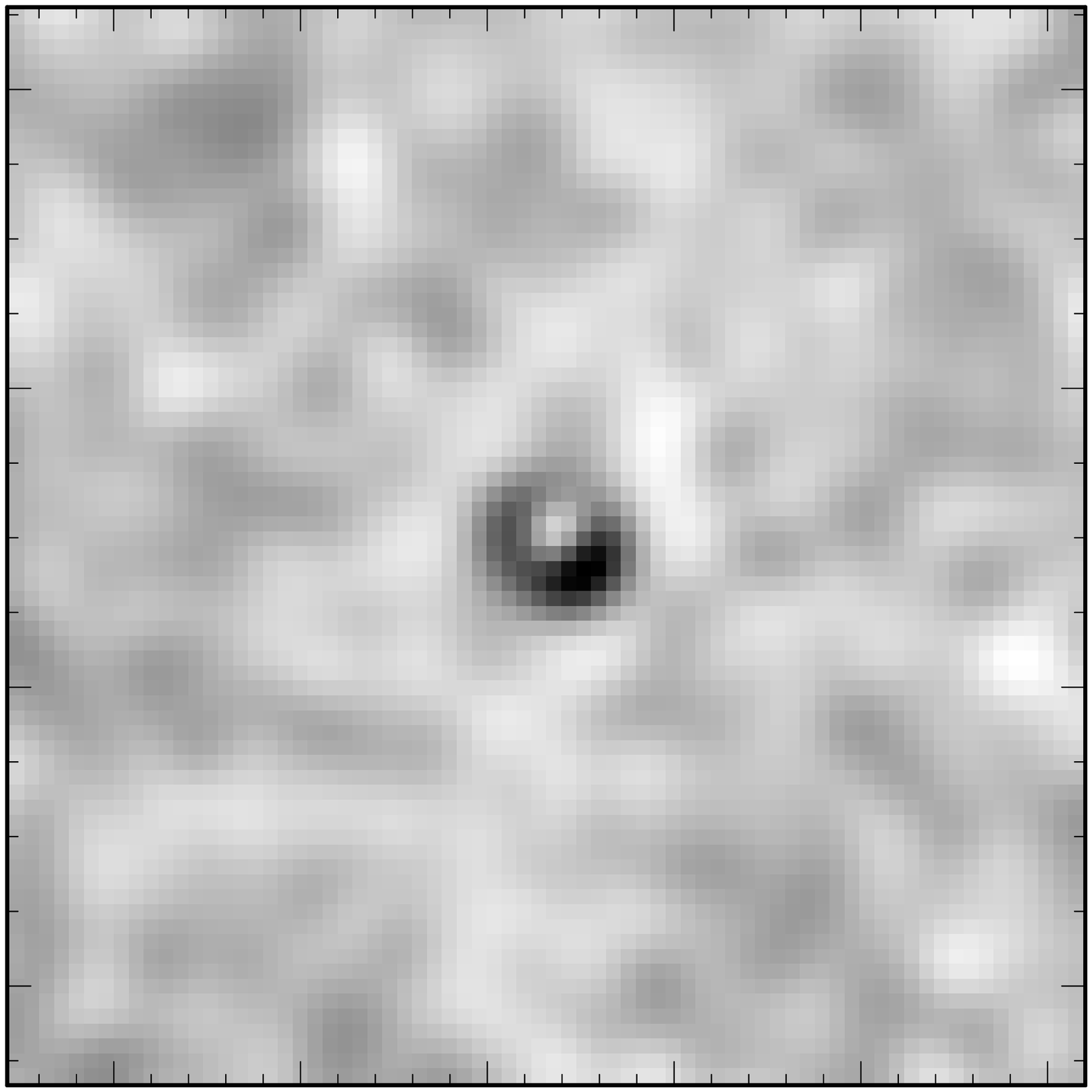} \\
\sex{} & Selavy\\
\includegraphics[width=0.4\linewidth]{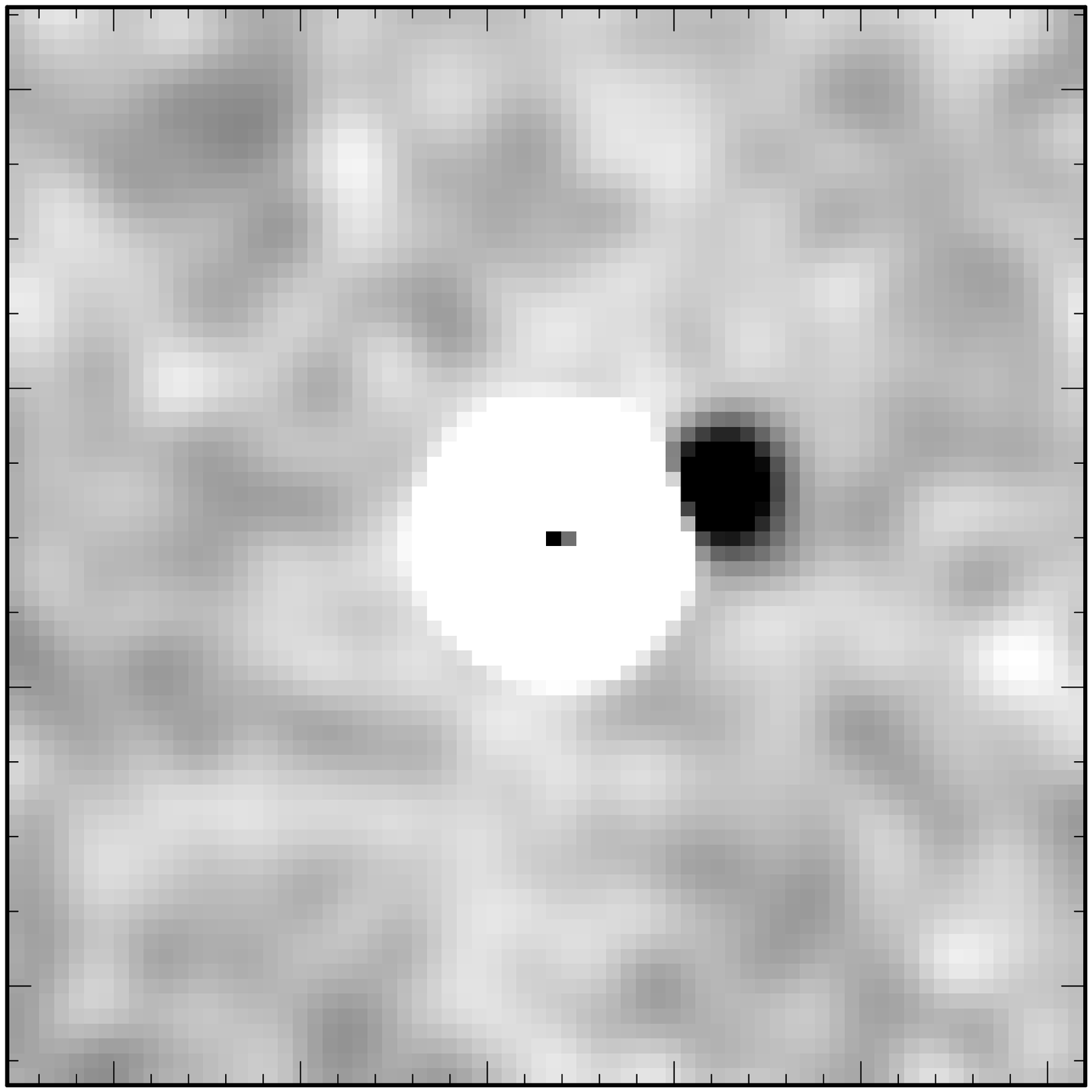} & 
\includegraphics[width=0.4\linewidth]{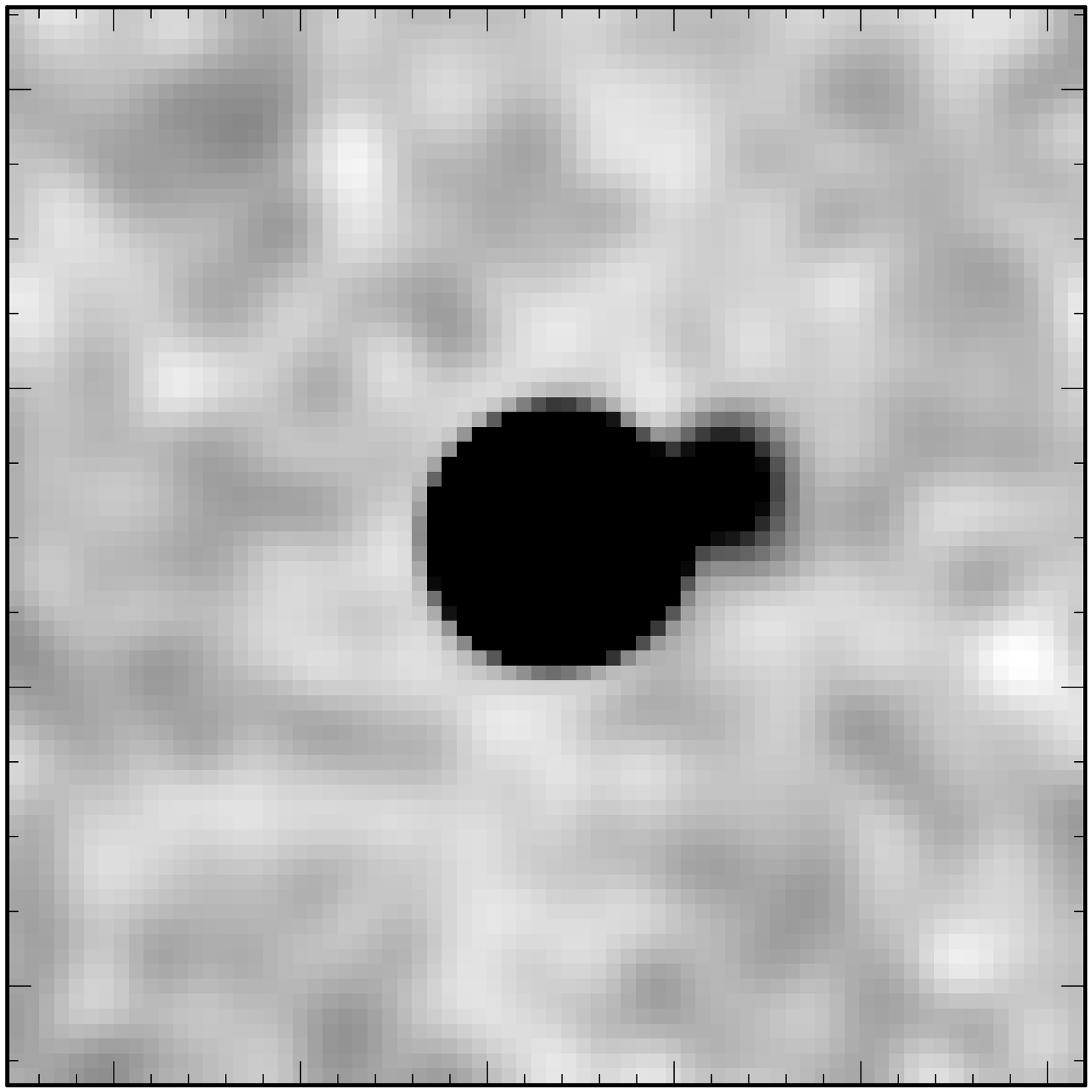} \\
\imsad{} & \tesla{} \\
\includegraphics[width=0.4\linewidth]{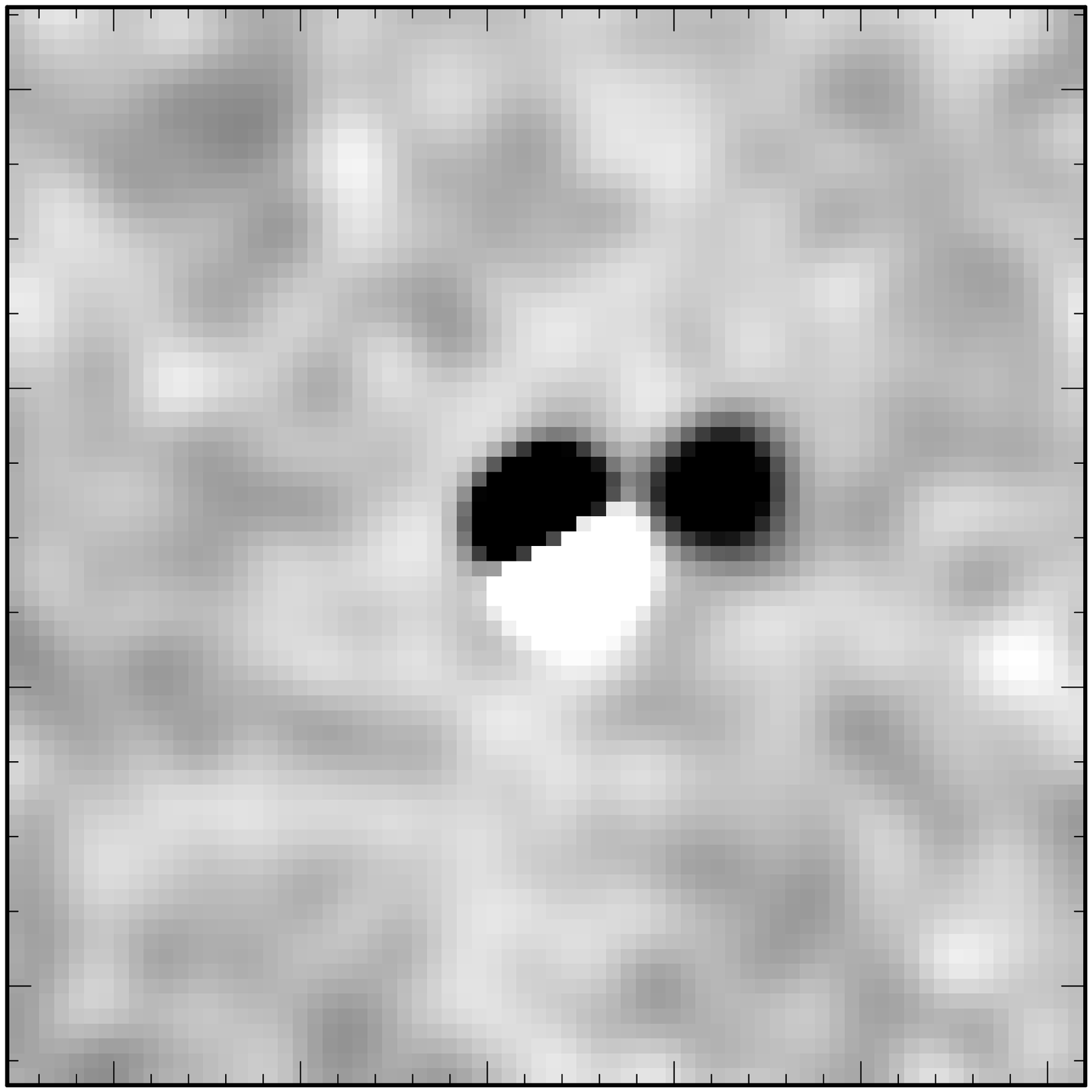} &
\includegraphics[width=0.4\linewidth]{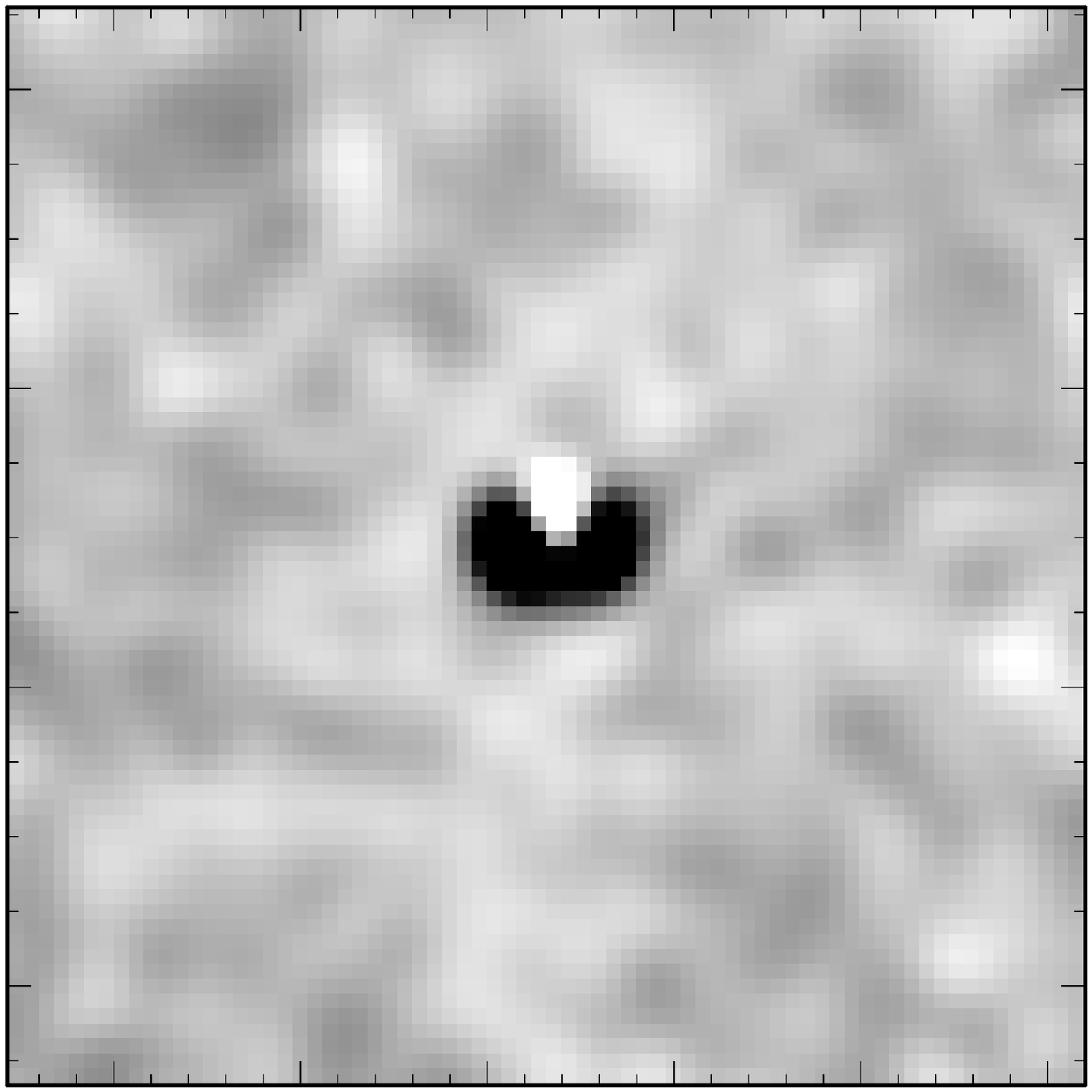} \\
\end{tabular}
\caption{{\em Top Left}: A section of the simulated image. {\em
Remainder}: The fitting residual for each of the source finding
algorithms. \tesla{} and \sfind{} were able to correctly identify and
characterise the two components but others were not.}
\label{fig:residual_90649}
\end{figure}

The second population of sources that is not well detected by the
source finding packages are the sources that are within an island of
pixels that contains multiple components. Examples of such islands are
shown in Figures \ref{fig:residual_196}, \ref{fig:residual_90649}, and
\ref{fig:fitresid}. If a source finding algorithm is unable to
correctly characterise multiple sources within an island, some or all
of these sources will be missed. There are two approaches used by the
tested algorithms to extract multiple sources from an island of pixels
- iterative fitting and de--blending. Each of these approaches can
fail to characterise an island of sources for different reasons, and
will now be discussed in detail.

\subsubsection{Iterative fitting}\label{sec:iterative}
The first approach to characterising an island of multiple components
is an iterative one which relies on the notion of a fitting
residual. The fitting residual is the difference between the data and
the model fit. In the iterative approach a single Gaussian is fit to
the island and the fitting residual is inspected. If the fitting
residual meets some criterion then the fit is considered to be `good'
and a single source is reported. If the residual is `poor' then the
fit is redone with an extra component. Once either the fitting
residual is found to be `good' or some maximum number of components
has been fit, the iteration stops and the extracted sources are
reported. A disadvantage of this method is that if the number of
allowed Gaussians ($n$) is poorly chosen, islands containing single
faint sources can have a `better' fitting residual when fit by
multiple components, and source fragmentation occurs. When a source is
fragmented it is difficult to extract the overall source parameters
from the multiple Gaussians that were used in the fitting of the
source. In particular the source flux is not simply the sum of the
flux of the fragments. If the chosen value of $n$ is too small then
not all of the sources within an island will be characterised.  These
uncharacterised sources will contaminate the fitting of the previously
identified sources resulting in a poor characterisation of the island.

\begin{figure}
\centering
\includegraphics[width=0.48\linewidth]{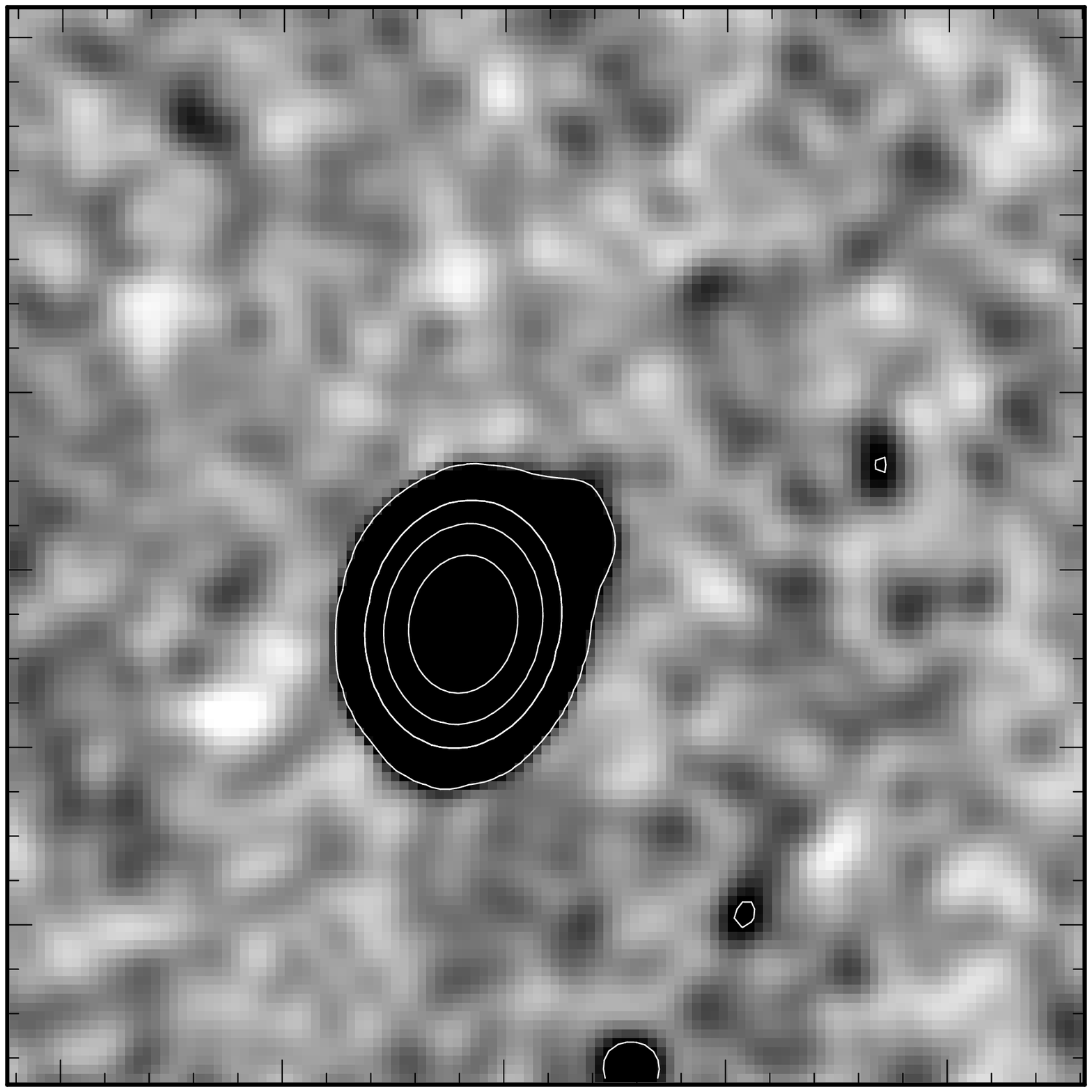}
\includegraphics[width=0.48\linewidth]{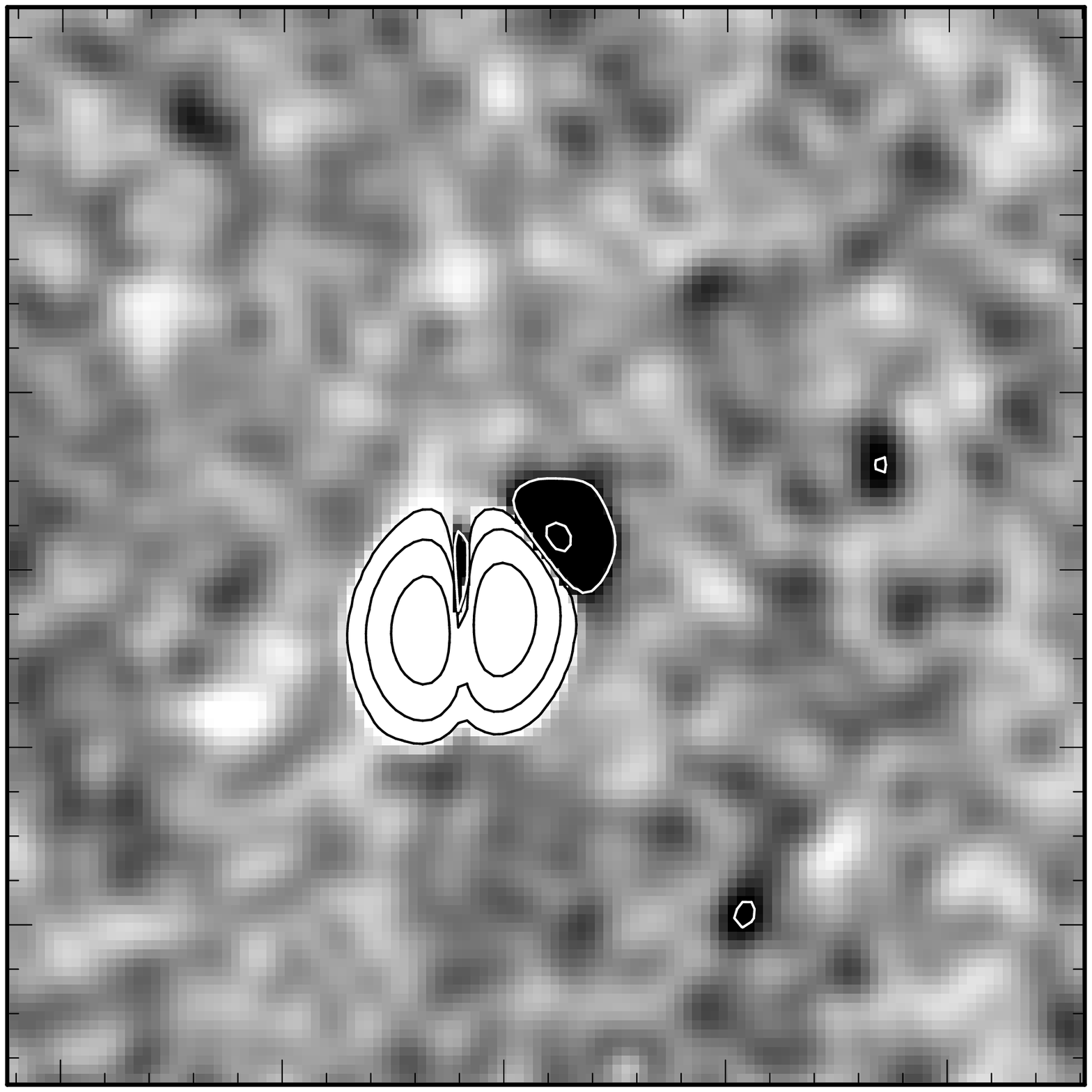}
\caption{{\em Left}: An island of pixels from the simulated image
containing both a $9\Jy$ source and a $1.7\mJy$ source. {\em Right}:
The fitting residual formed by subtracting a (\tesla{}) fitted model of
the $9\Jy$ source from the data. The pixel scale is $-3\sigma$ (white)
to $+5\sigma$ (black) with contours at an SNR of $\pm5, \pm 50, \pm
500, \pm5000$ in contrasting tones. The flux of the source and its
major axis have both been measured to within $0.05\%$ of the true
value and yet the fitting residual has peaks at an SNR of over 500.}
\label{fig:fitresid}
\end{figure}

When the flux ratio of components within an island of pixels becomes
very large, an iterative fitting approach can fail. The cause of this
failure is related to the performance of an ideal Gaussian fitting
routine. Figure~\ref{fig:fluxes} shows the fractional error in
measuring the amplitude of a Gaussian. For high SNR
sources, the absolute flux error can be orders of magnitude below the
rms image noise, so it may be expected that the maximum flux in the
fitting residual should also be at or below the rms image
noise. However the main contribution to the flux seen in the fitting
residual is not from amplitude errors but from errors in estimating
the FWHM of the source.

The amplitude difference between a (1D) Gaussian of amplitude A and
FHWM of $\theta$ ($=2\sqrt{2\log{2}}\sigma$) and a second Gaussian of
identical amplitude A and FWHM of $\theta^\prime =
\theta+\Delta\theta$, is given by $F(x)$:
\begin{align}
F(x) &= A\left( e^{-\frac{x^2 4\ln2}{\theta^2}} - e^{-\frac{x^2 4\ln2}{{\theta^\prime}^2}} \right),
\end{align}
which has maxima at
\begin{align}
x_0^2 &= \frac{\ln\frac{\theta}{\theta^\prime}}{2\ln2}
\left(\frac{{\theta^\prime}^2\theta^2}{{\theta^\prime}^2-\theta^2}
\right).
\end{align}
As a fraction of the true flux, the maximum residual is then
\begin{align}
\frac{F(x_0)}{A} &= \left( \frac{2\Delta\theta}{\theta}\right) \times
\left( 1+\frac{2\Delta\theta}{\theta}\right)^{\frac{\theta}{2\Delta\theta}} 
\end{align}
The typical error in the measurement of $\theta$ is \citep{Condon1997}
\begin{align}
\frac{\Delta\theta}{\theta}&=\frac{\mu(\theta)}{\theta} \simeq \frac{\sigma}{A},
\end{align}
so that a source with an SNR of $A/\sigma$ will have a fitting residual with an SNR of
\begin{align}
\frac{F(x_0)}{\sigma} &= 2\left( 1+\frac{2\sigma}{A}\right)^{\frac{A}{2\sigma}}. \label{eq:residual}
\end{align}
From the Equation~\ref{eq:residual} it is clear that in an island
whose brightest source has an SNR of $A/\sigma$, sources below an SNR
of $F(x_0)/\sigma$ will not be detected by an iterative fitting
method. The fitting residual exceeds $5\sigma$ at an SNR as low as
11. Therefore, even an ideal Gaussian fitting routine will miss
$5\sigma$ sources that are within the same island as a source of $\geq
11\sigma$ if an iterative approach is taken. If two Gaussian
components are fit to an island of pixels such as that shown in
Figure~\ref{fig:fitresid}, and the positions are left unconstrained,
the fainter component will migrate towards one of the maxima in the
fitting residual. The brighter source will then be characterised by
two Gaussians, and the fainter source by none. The final result, is
that neither of the sources will be well characterised. It is
therefore essential that a source finding algorithm has some method
for determining the number of Gaussian components within an island, as
well as a way to stop the fitting process from mis-characterising the
two sources. A process called sectioning or de-blending is a common
method.

\subsubsection{Sectioning or de-blending}
A second approach to characterising islands with multiple sources is
to use the distribution of flux within the island to determine the
number of components to be fit, and then fit the components. This
approach relies on some {\em a priori\/} knowledge of what a source
looks like to break an island into components. \sfind{}, \sex{}, and
\tesla{} all use a form of sectioning to generate an initial estimate
of the number of sources to be fit, as well as the starting
parameters.

It is possible to create a statistical measure that will account for
the number of sources that are missed because there are multiple
sources within an island of pixels. This would, however, require
detailed knowledge of the source finding algorithm, the flux
distribution of the source population, and the flux dependent
two-point correlation function. The complexity of this calculation
means that it is never computed and sometimes not even
considered. Since many variable phenomena appear in or near known
sources (eg., radio supernovae in galaxies, extreme scattering events
within our own Galaxy, and more), an inability to accurately
characterise this population of sources will make it difficult
or impossible to reliably detect and characterise many variable
events.

\section{The new source finding program: \tesla{}} \label{sec:tesla}
With an understanding of how the underlying algorithms affect a source
finder's ability to find and accurately characterise islands of
pixels, we have created a new source finding algorithm. The goal of
the new algorithm is to incorporate the reliability and completeness
performance of the packages studied in
\S~\ref{sec:alg}-\ref{sec:missed}, whilst improving on their ability
to characterise islands of pixels. The source finding algorithm is
called \tesla{}, as it deals with many islands.

As background estimation and subtraction are not part of the focus of
this work, \tesla{} has been designed with only a simple background
estimation algorithm. For the analysis presented, \tesla{} was run
with a detection threshold of $125\uJy/\rm{beam}$. \tesla{} uses the
FloodFill algorithm described in \S~\ref{sec:floodfill} to create
islands of pixels.

\tesla{} makes use of the notion of a single curvature map to
characterise an island of pixels. The curvature $\kappa$ of a function
$f(x)$ is given by:
\begin{align}
  \kappa &= \frac{f^{\prime\prime}}{(1+f^{\prime^2})^{3/2}}
\end{align}
\citep{reilly_mean_1982}. For a Gaussian with a FHWM of $k$ pixels,
\begin{align}
  f^{\prime}(x) &= \frac{-16x\ln2}{k^2} e^{-\frac{x^28\ln2}{k^2}},
\end{align}
so that $f^{\prime^2}$ has a maxima at $x=k/\sqrt{2}$, and 
\begin{align}
  f^{\prime^2} &\leq \frac{1}{k^2} \frac{(\ln2)^2}{2^9}.
\end{align}
For a Gaussian with $k\geq 1$, $ f^{\prime^2}\ll 1$ and we can approximate
\begin{align}
  \kappa &\simeq f^{\prime\prime}.\label{eq:curvature}
\end{align}
The curvature of a surface in a particular direction can be defined
using Equation~\ref{eq:curvature}, where the differentiation is along
a unit vector in the chosen direction. \citet{molinari_source_2011}
calculate the curvature of their input image in four image directions
in order assist their source finding and characterisation. We combine
these four curvature measurements to calculate the mean curvature of
an image. For an image convolved with a Gaussian with a FWHM of $k$
pixels, the (mean) curvature, $\bar\kappa$ is equal to the mean of
$\kappa$ calculated in any two orthogonal directions
\citep{reilly_mean_1982}. The discrete 2D Laplacian kernel
\begin{equation}
L^2_{xy} =  \left[ \begin{array}{ccc}
1 & 1 & 1  \\
1 & -8 & 1 \\
1 & 1 & 1 \end{array} \right],
\end{equation}
calculates the sum of the second derivatives in four
directions. Convolving the input image with $L^2_{xy}$ will therefore
produce a map of $2\bar\kappa$ - a single curvature map.

\begin{figure}
\centering
\includegraphics[width=\linewidth]{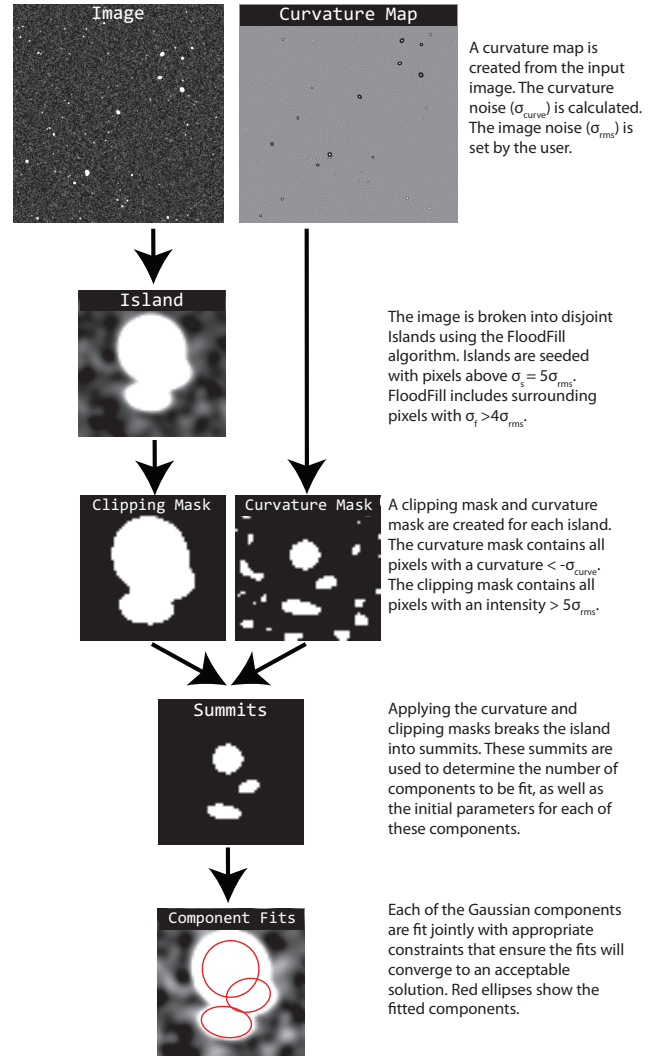}
\caption{A demonstration of the operation of \tesla{}, using a single
multiple component island as example. The input image is used to
create a curvature map from which the curvature noise $\sigma_{curve}$
is calculated. FloodFill is used to break the image into islands of
pixels. The number of components in an island is estimated using a
combination of the curvature mask and threshold mask. An elliptical
Gaussian is fit to each of the components in the island
simultaneously.}
\label{fig:tesla_alg}
\end{figure}

\begin{figure}
\centering
\includegraphics[width=0.45\linewidth]{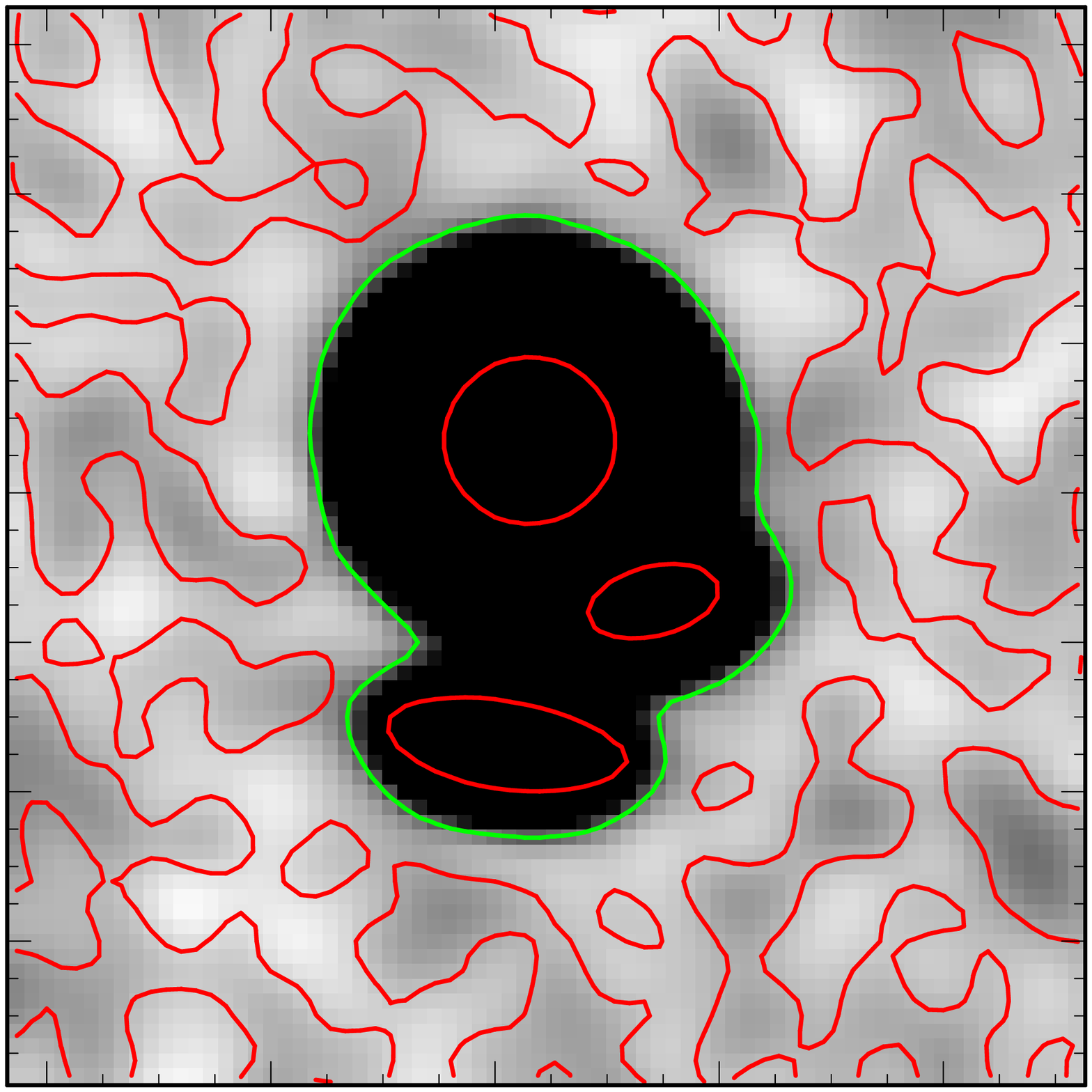}
\includegraphics[width=0.45\linewidth]{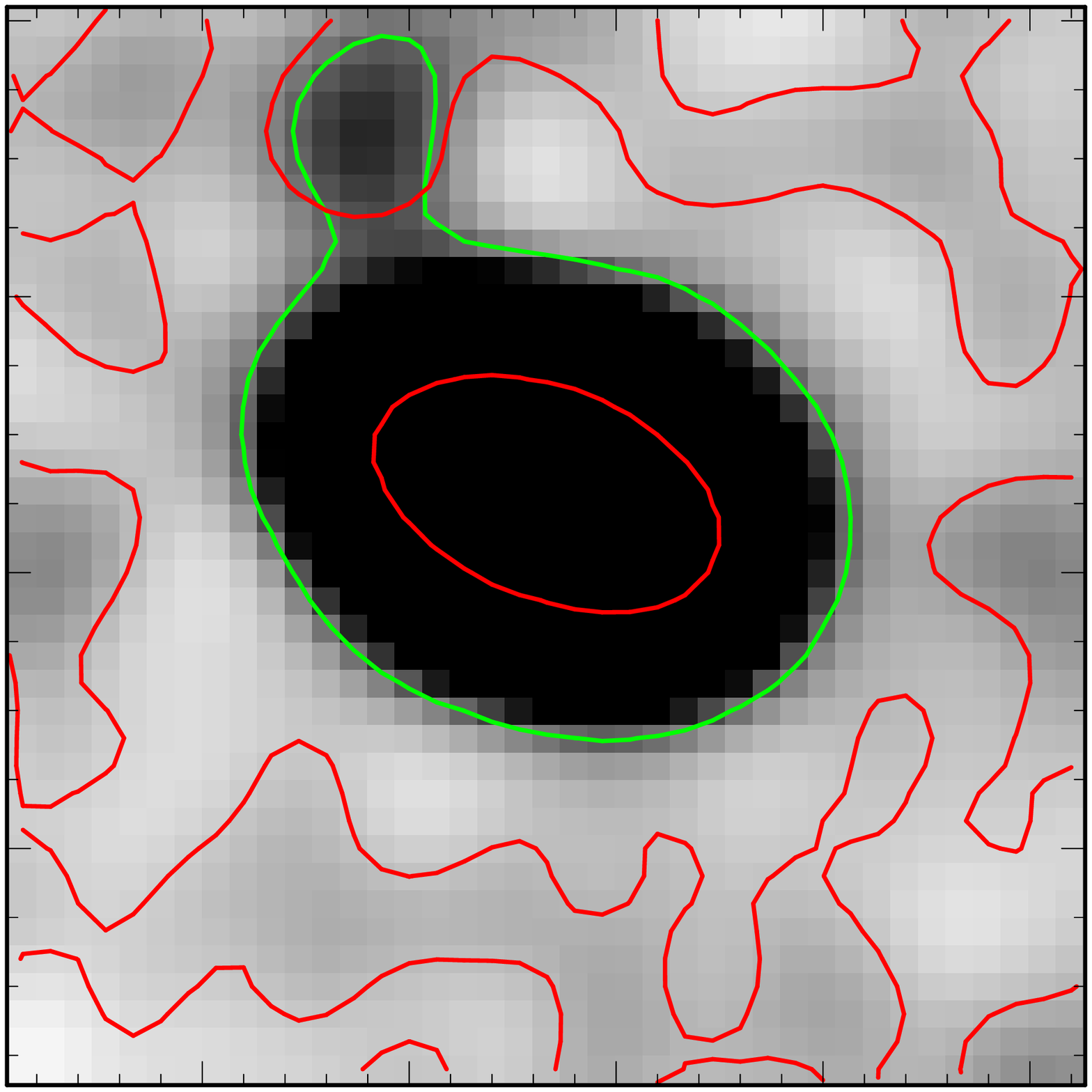}
\caption{Two examples of the curvature analysis scheme. The greyscale
represents the flux density map and ranges from $-3\sigma$ (white) to
$+10\sigma$ (black). The green contour is at $5\sigma$ and represents
the island boundary.  The red contours are where the curvature map
changes from positive to negative. Regions surrounded by a red contour
have negative curvature and are the local maxima.}
\label{fig:curvature}
\end{figure}

Islands of pixels are fit with multiple Gaussian components. The
number of components to be fit is determined from a curvature map.
The curvature map will be negative around local maxima. Groups of
contiguous pixels that have negative curvature and fluxes above the
threshold are called summits. An island of pixels will contain one or
more summits. \tesla{} fits one component per summit, with the
parameters of each of the components are taken from the corresponding
summit. The position and flux are initially set to be equal to the
brightest pixel within a summit, and the shape parameters (major/minor
axis and position angle) are set to be the same as the convolving
beam. Figure~\ref{fig:curvature} shows an example of two islands that
contain multiple sources with the island boundaries and regions of
negative curvature delimited. In the example in the left panel of
Figure~\ref{fig:curvature} there are three regions of negative
curvature that are completely within the green island. This island is
fit with three Gaussians. In the example in the right panel of
Figure~\ref{fig:curvature} there are two regions of negative curvature
that overlap with the island of pixels. One component is contained
entirely within the island, whilst the other is only partly within the
island. Only the region of negative curvature that is within the green
island is considered when estimating the initial parameters of the
components. Both of the islands depicted in Figure~\ref{fig:curvature}
contains a source that is bright enough that the expected fitting
residual would be brighter than any of the other components within the
island, and therefore an iterative fitting approach would only fit a
single component (see \S~\ref{sec:iterative}). Since the island of
pixels in the right panel of Figure~\ref{fig:curvature} has two
summits, \tesla{} is able to accurately detect and characterise both
components.  Islands of pixels that contain only a single source have
only a single summit and are fit with a single component.

To avoid faint components migrating to the fitting residual of
brighter components, the position of each of the components is
constrained to be within the corresponding summit. The flux of each
component must be greater than $5\sigma$. For low SNR sources, the
true flux can be significantly different from the intensity of the
brightest pixel in the summit, $S_{max}$. For high SNR sources such
noise variations are less important and beam sampling effects become
more important.

For an image with a sampling rate of $k$ pixels per beam a source of
flux $S$ which is located at the intersection of four pixels will
effectively be sampled $\sqrt{2}*\theta/2k$ pixels from the centre of
the source. The intensity of the peak pixel is therefore given by:
\begin{align}
S_{max}&=S\cdot\exp\left(- \frac{\left( \frac{\sqrt{2}\theta}{2k}\right)^2 *4\ln{2}}{\theta^2} \right)\\
 &= S\cdot2^{-\frac{2}{k^2}},
\end{align}
where $\theta$ is the FWHM of the source. The flux of each component
is therefore constrained to be less than
$S_{max}\cdot2^{\frac{2}{k^2}} +3\sigma$.

A Gaussian function has negative curvature from the peak out to
$\pm{\rm FWHM}/\sqrt{2}$. The size of a summit is therefore used to
constrain the component size. The major and minor axes of a component
must be larger than the synthesised beam, and must remain smaller than
$\sqrt{2}$ times the width of the summit. Beam sampling effects again
play a role here, and so in \tesla{}, we increase the limits on the major
and minor axes each by two pixels to account for this. If the summit
is smaller than the synthesised beam then the component is fit with
the PSF.

The performance of \tesla{} has been presented in
\S~\ref{sec:eval}-\ref{sec:missed} along with the other source
finding algorithms under study.

\section{Conclusions} \label{sec:conclusions}
Using a simulated data set, we have assessed the performance of some
widely used source finding packages, along with the ASKAPsoft source
finding program Selavy. These source finding packages are found to
produce complete and reliable catalogues of isolated compact
sources. We identify two populations of sources that are not well
detected by the source finding packages. The first population being
faint sources close to the detection limit, and the second being
sources which are within an island of pixels containing multiple
components. Islands of pixels with multiple components are found to be
poorly characterised by source finding packages that take an iterative
fitting approach to characterisation. Source finding packages that
estimate the number of components in an island prior to fitting are
less likely to mis-characterise the island. We have developed a new
source finding package, \tesla{}, which is able to characterise the
number of components within an island of pixels more accurately than
any of the other packages tested.

\tesla{} makes use of a curvature image which is derived from the input image
with a Laplacian transform. Using the curvature image \tesla{} is able to
accurately determine the number of compact components within an island
of pixels and produce a set of initial parameters and limits for a
constrained fit of multiple elliptical Gaussians.

\tesla{} has been shown to produce catalogues with a $5\sigma$
completeness that is better than our estimation of an ideal source
finder. This completeness has been achieved without sacrificing
reliability, and \tesla{} is the most reliable of the tested
algorithms. The next generation of radio surveys will be sensitive
enough that $\sim5\%$ of the islands in the image will contain
multiple components and therefore the ability to characterise such
islands is of critical importance. \tesla{} is able to accurately
characterise islands of pixels which contain multiple compact
components. 

We have shown that in order to improve the reliability and
completeness of source catalogues it is necessary to perform
constrained multiple Gaussian fitting. An accurate estimation of
initial parameters and sensible constraints are both critical when
multiple component Gaussian fitting is performed.  We have
demonstrated a method for estimating and constraining the fitting
parameters which is based on the curvature of the image. We anticipate
that by adopting the \tesla{} algorithm, the next generation of radio
continuum surveys will be able to achieve more complete, reliable and
accurate catalogues without relying on significant manual
intervention.

\section*{Acknowledgments}
The authors thank David McConnell, David Kaplan, and Theodora
Papadimatos for useful discussions about source finding
algorithms. TM, BMG and PH acknowledge support from the Australian
Research Council through Super Science Fellowship grant FS100100033
and JRC through Discovery Projects grant DP1097291. The Centre for
All--sky Astrophysics is an Australian Research Council Centre of
Excellence, funded by grant CE11E0090.

\bibliographystyle{apj}
\bibliography{ms}

\begin{thebibliography}{20}
\expandafter\ifx\csname natexlab\endcsname\relax\def\natexlab#1{#1}\fi

\bibitem[{Adams {et~al.}(2004)Adams, Bunton, \& Kesteven}]{adams_square_2004}
Adams, T.~J., Bunton, J.~D., \& Kesteven, M.~J. 2004, ExA, 17, 279

\bibitem[{Bannister {et~al.}(2011)Bannister, Murphy, Gaensler, Hunstead, \&
  Chatterjee}]{bannister_2011}
Bannister, K.~W., Murphy, T., Gaensler, B.~M., Hunstead, R.~W., \& Chatterjee,
  S. 2011, MNRAS, 412, 634

\bibitem[{Becker {et~al.}(1995)Becker, White, \& Helfand}]{becker_first_1995}
Becker, R.~H., White, R.~L., \& Helfand, D.~J. 1995, ApJ, 450, 559

\bibitem[{Bertin \& Arnouts(1996)}]{Bertin1996}
Bertin, E., \& Arnouts, S. 1996, A\&AS, 117, 393

\bibitem[{Chatterjee {et~al.}(2010)Chatterjee, Murphy, \& {VAST
  Collaboration}}]{chatterjee_vast_2010}
Chatterjee, S., Murphy, T., \& {VAST Collaboration}. 2010, in Bulletin of the
  American Astronomical Society, Vol.~42, American Astronomical Society Meeting
  Abstracts \#215, \#470.12--+

\bibitem[{Condon(1997)}]{Condon1997}
Condon, J.~J. 1997, PASP, 109, 166

\bibitem[{Condon {et~al.}(1998)Condon, Cotton, Greisen, Yin, Perley, Taylor, \&
  Broderick}]{condon_nrao_1998}
Condon, J.~J., Cotton, W.~D., Greisen, E.~W., Yin, Q.~F., Perley, R.~A.,
  Taylor, G.~B., \& Broderick, J.~J. 1998, Astronomical Journal, 115, 1693

\bibitem[{Croft {et~al.}(2011)Croft, Bower, Keating, Law, Whysong, Williams, \&
  Wright}]{croft_2011}
Croft, S., Bower, G.~C., Keating, G., Law, C., Whysong, D., Williams, P. K.~G.,
  \& Wright, M. 2011, ApJ, 731, 34

\bibitem[{Eddington(1913)}]{eddington_1913}
Eddington, A.~S. 1913, MNRAS, 73

\bibitem[{Hopkins {et~al.}(2002)Hopkins, Miller, Connolly, Genovese, Nichol, \&
  Wasserman}]{hopkins2002}
Hopkins, A.~M., Miller, C.~J., Connolly, A.~J., Genovese, C., Nichol, R.~C., \&
  Wasserman, L. 2002, AJ, 123, 1086

\bibitem[{Huynh {et~al.}(2011)Huynh, Hopkins, Norris, Hancock, Murphy, \&
  Jurek}]{huynh_emu_2011}
Huynh, M.~T., Hopkins, A.~M., Norris, R.~P., Hancock, P.~J., Murphy, T., \&
  Jurek, R. 2011, PASA (in press)

\bibitem[{Johnston {et~al.}(2008)Johnston, Taylor, Bailes, Bartel, Baugh,
  Bietenholz, Blake, Braun, Brown, Chatterjee, Darling, Deller, Dodson,
  Edwards, Ekers, Ellingsen, Feain, Gaensler, Haverkorn, Hobbs, Hopkins,
  Jackson, James, Joncas, Kaspi, Kilborn, Koribalski, Kothes, Landecker, Lenc,
  Lovell, Macquart, Manchester, Matthews, McClure-Griffiths, Norris, Pen,
  Phillips, Power, Protheroe, Sadler, Schmidt, Stairs, Staveley-Smith, Stil,
  Tingay, Tzioumis, Walker, Wall, \& Wolleben}]{johnston_science_2008}
Johnston, S., {et~al.} 2008, ExA, 22, 151

\bibitem[{Lonsdale {et~al.}(2009)Lonsdale, Cappallo, Morales, Briggs,
  Benkevitch, Bowman, Bunton, Burns, Corey, DeSouza, Doeleman, Derome,
  Deshpande, Gopala, Greenhill, Herne, Hewitt, Kamini, Kasper, Kincaid, Kocz,
  Kowald, Kratzenberg, Kumar, Lynch, Madhavi, Matejek, Mitchell, Morgan,
  Oberoi, Ord, Pathikulangara, Prabu, Rogers, Roshi, Salah, Sault, Shankar,
  Srivani, Stevens, Tingay, Vaccarella, Waterson, Wayth, Webster, Whitney,
  Williams, \& Williams}]{lonsdale_mwa_2009}
Lonsdale, C.~J., {et~al.} 2009, Proceedings of the IEEE, 97, 1497

\bibitem[{Mauch {et~al.}(2003)Mauch, Murphy, Buttery, Curran, Hunstead,
  Piestrzynski, Robertson, \& Sadler}]{mauch2003}
Mauch, T., Murphy, T., Buttery, H.~J., Curran, J., Hunstead, R.~W.,
  Piestrzynski, B., Robertson, J.~G., \& Sadler, E.~M. 2003, MNRAS, 342, 1117

\bibitem[{Molinari {et~al.}(2011)Molinari, Schisano, Faustini, Pestalozzi, {Di
  Giorgio}, \& Liu}]{molinari_source_2011}
Molinari, S., Schisano, E., Faustini, F., Pestalozzi, M., {Di Giorgio}, A.~M.,
  \& Liu, S. 2011, Astronomy \& Astrophysics, 530, A133

\bibitem[{Murphy {et~al.}(2007)Murphy, Mauch, Green, Hunstead, Piestrzynska,
  Kels, \& Sztajer}]{murphy_second_2007}
Murphy, T., Mauch, T., Green, A., Hunstead, R.~W., Piestrzynska, B., Kels,
  A.~P., \& Sztajer, P. 2007, MNRAS, 382, 382

\bibitem[{Norris {et~al.}(2011)Norris, Hopkins, Afonso, Brown, Condon, Dunne,
  Feain, Hollow, Jarvis, Johnston-Hollitt, Lenc, Middelberg, Padovani,
  Prandoni, Rudnick, Seymour, Umana, Andernach, Alexander, Appleton, Bacon,
  Banfield, Becker, Brown, Ciliegi, Jackson, Eales, Edge, Gaensler, Giovannini,
  Hales, Hancock, Huynh, Ibar, Ivison, Kennicutt, Kimball, Koekemoer,
  Koribalski, L\'{o}pez-S\'{a}nchez, Mao, Murphy, Messias, Pimbblet,
  Raccanelli, Randall, Reiprich, Roseboom, R\"{o}ttgering, Saikia, Sharp, Slee,
  Smail, Thompson, Urquhart, Wall, \& Zhao}]{norris_emu_2011}
Norris, R.~P., {et~al.} 2011, PASA, 28, 215

\bibitem[{Reilly(1982)}]{reilly_mean_1982}
Reilly, R.~C. 1982, The American Mathematical Monthly, 89, 180

\bibitem[{Roerdink \& Meijster(2001)}]{roerdink_watershed_2001}
Roerdink, J., \& Meijster, A. 2001, Fundamenta Informaticae, 41, 187

\bibitem[{Sault {et~al.}(1995)Sault, Teuben, \&
  Wright}]{sault_retrospective_1995}
Sault, R.~J., Teuben, P.~J., \& Wright, M. C.~H. 1995, in Astronomical Society
  of the Pacific Conference Series, Vol.~77, Astrnonomical Data Analysis
  Software and Systems IV, ed. R.~Shaw, H.~Payne, \& J.~Hayes, 433

\end{thebibliography}
\label{lastpage}
\end{document}